%% file: ms_apj.tex
\documentclass[12pt,preprint]{emulateapj}
\usepackage{natbib,subfigure}
\newcommand{\myemail}{p.cargile@vanderbilt.edu}
\slugcomment{Accepted for Publication in the Astronomical Journal}
\shorttitle{BVI$_{C}$ Photometric Survey of IC~4665}
\shortauthors{Cargile \& James}
\begin{document}
\title{Employing a New BVI$_{c}$ Photometric Survey of IC~4665 to Investigate the Age of this Young Open Cluster}
\author{P.~A. Cargile\altaffilmark{1},  D.~J. James\altaffilmark{1,2}}
\altaffiltext{1}{Department of Physics and Astronomy, Vanderbilt University, Nashville, TN 37235, USA, \myemail}
\altaffiltext{2}{Physics and Astronomy Department, University of Hawai'i at Hilo, Hilo, HI 96720, USA}

\begin{abstract}
We present a new, BVI$_{c}$ photometric survey of the young open cluster IC~4665, which improves on previous studies of this young cluster by incorporating a rigorous standardization procedure, thus providing high-fidelity colors and magnitudes for cluster members. We use this new photometric dataset to reevaluate the properties (age and distance) of IC~4665. Namely, using a statistical approach incorporating $\tau^{2}$ CMD modeling, we measure a pre-main-sequence isochrone age and distance of 36$\pm$9 Myr and 360$\pm$12 pc, as well as a upper-main-sequence {\em turn-off} age and distance of 42$\pm$12 Myr and 357$\pm$12 pc. These ages and distances are highly dependent on the isochrone model and color used for the fitting procedure, with a possible range of $\sim$10-20 Myr in age and $\sim$20 parsecs in distance. This spread in calculated ages and distances seen between colors and models is likely due to limitations in the individual membership catalogs and/or systematic differences in the predicted stellar parameters from the different sets of models. Interestingly, when we compare the isochrone ages for IC~4665 to the published lithium depletion boundary age, 28$\pm$5 Myr, we observe that this cluster does not appear to follow the trend of isochrone ages being 1.5 times smaller than lithium depletion boundary ages. In addition, comparing the overall magnetic activity (X-ray and H$\alpha$ emission) in IC~4665 with other well studied open clusters, we find the observed activity distributions for this young cluster are best characterized by assuming an age of 30--40 Myr, thus in agreement with our pre-main-sequence and {\em turn-off} isochrone ages for IC~4665. Overall, although some age discrepancies do exist, particularly in the ages measured from pre-main-sequence isochrones, the range of possible IC~4665 ages derived from the various dating techniques employed here is relatively small compared to that found for other well studied open clusters.
\end{abstract}
\section{Introduction}\label{sec.intro}

Open clusters have long provided necessary empirical constraints for our current models of stellar evolution due to their large collection of coeval, equidistant, and chemically homogeneous stars. Through observations of large numbers of open clusters with known ages, we are able to investigate the manner with which fundamental stellar properties evolve as stellar evolution progresses. This is particularly true for early stellar evolution (stellar ages $<$ 100 Myr), when fundamental physical parameters of stars change rapidly ({\em e.g.,} radius, temperature, rotation rate, magnetic field strength). However, for these clusters to be used as valuable fiducial benchmarks of stellar evolution, we must be able to date them in a consistent manner with high-fidelity and precision. Currently, there are several dating methods utilized to determine open cluster ages, including fitting theoretical model isochrones to empirical H-R diagrams, light element abundance distributions, and exploiting the connection between magnetic activity and age.

A heavily relied upon approach to determining absolute ages of star clusters is finding a best-fitting color-magnitude diagram (CMD) isochrone for the nuclear turn-off \citep{Sandage1958,Mermilliod1981a,Mermilliod1981b,Meynet1993} and/or the pre-main-sequence (PMS) turn-on \citep{DAntona1997,Baraffe1998,Siess2000} to the cluster's main-sequence, typically performed using an {\em ad-hoc} `by-eye' comparison of the model isochrone to the observational photometric data. This technique, however easily applied in theory, is in practice commonly hampered by uncertainties including field star contamination, cluster distance and/or photometric error. Furthermore, this technique suffers from the measured cluster ages being heavily dependent on the input physics used by the stellar evolution models. For instance, one can show for the same cluster that different nuclear turn-off and PMS models can give stellar ages that differ by a factor of 1.5-2 \citep{Chiosi1992,Meynet1993,Meynet1997,Meynet2000,Baraffe1998,Baraffe2002,Siess2000,Naylor2009}.

There now exist several alternative dating techniques that employ other observable properties of stars to find absolute and relative ages for open clusters. Observations have shown that the level of magnetic activity, indicated by the presence of X-ray and/or chromospheric activity, in solar-type stars clearly decreases with increasing age \citep{Wilson1963,Skumanich1972,Noyes1984,Soderblom1991,Mamajek2008,Cargile2009}. This relationship is fundamentally tied to the interplay between magnetic activity and the rotationally-induced, magneto-hydrodynamic dynamo in solar-type stars. This causal relationship, the so called age-rotation-activity paradigm, generally holds true; however, it can be affected by natural activity cycles, similar to those seen in the Sun. For instance, \citet{Soderblom1991} show that uncertainties in stellar chromospheric ages can be as large as $50\%$ for individual stars. Of course, in open clusters this error can be statistically reduced by measuring many stars. 

IC~4665 is currently one of only a handful of well-studied PMS open clusters ({\em i.e.,} age of 5-50 Myr), {\em e.g.,} NGC~2169 \citep[10 Myr][]{Jeffries2007b}, NGC~2547 \citep[35 Myr][]{Jeffries2005}, and IC~2391 \citep[50 Myr][]{BarradoyNavascues2004}. IC~4665 thus offers an interesting, additional opportunity to place empirical constraints on models of stellar evolution at young ages. Because of this, IC~4665 has been surveyed extensively to determine cluster membership, as well as observe physical properties of cluster members, including studies on the kinematics of the cluster \citep{Prosser1993,Prosser1994,Manzi2008,Jeffries2009}, a ROSAT survey of stellar X-ray emission \citep{Giampapa1998}, and studies of the distribution of Li abundances in the cluster \citep{Martin1997,Manzi2008,Jeffries2009}. Most of these earlier investigations into the properties of IC~4665 members relied on the optical photometry and membership catalogs reported in \citet[P93]{Prosser1993}. Unfortunately, the cluster members detailed in this study rely upon uncertain photometric data, due to the use of an improper photometric standardization procedure (see Section~\ref{sec.comp}), as well as having a high level of field star contamination as a result of their membership catalog being based upon unreliable proper motions (see Section \ref{sec.lowmass}).

In light of such uncertainties, we report a new, standardized BVI$_{c}$ photometric survey of the central 1 sq.~deg.~of the cluster (see Section \ref{sec.newphot}). We use this new catalog to re-derive the age of IC~4665 using several different methods, including modeling the PMS (Section \ref{sec.lowmass}) and upper-main-sequence (Section \ref{sec.highmass}) photometric membership, and comparing the X-ray emission (Section \ref{sec.xray}) and H$\alpha$ emission (Section \ref{sec.halpha}) distributions to other well known open clusters. In addition, we compare our new age measurements in the context of the available ages for IC~4665 in the literature.

\section{The Open Cluster IC~4665}\label{sec.overview}

IC~4665 is a relatively young ($<$ 100 Myr), nearby ($\sim$350 pc) open cluster in the constellation Ophiuchus (R.A.(2000): 17$^{h}$46$^{m}$, Dec(2000):+05$\arcdeg$43$^{'}$). The WEBDA Open Cluster Database\footnote{The WEBDA database, developed by J.-C. Mermilliod, can be found at http://www.univie.ac.at/webda/} reports for the cluster an age of 43 Myr, a distance of 365 pc, a high Galactic latitude ($b = +17 \deg$), and has a reddening of $E(B-V)=0.174$ (values originally published in \citealt{Dias2002}). The youth and close proximity of this cluster has made it a popular target for studying stars during early times, while the majority of its low-mass members are still in the PMS phase.

\subsection{Extant Observations of IC~4665}\label{sec.ext}

Previously, the most extensive optical survey of IC~4665 was performed by \citet{Prosser1993}, which included astrometric, photometric, and spectroscopic programs to identify members of the cluster. P93 acquired BVRI$_{k}$ CCD photometry for 4100 stars down to a limiting magnitude of $V\sim18.5$. Proper motions were measured from archival photographic plates, and membership probabilities were derived for stars brighter than V$\simeq$14.5. Their analysis found 170 stars with non-zero proper motion membership probability, of which 22 were identified as previously known cluster members, 57 were classified as photometric candidates ({\em i.e.,} stars lying near a 36 Myr isochrone), 72 photometric nonmembers, and 22 stars left with undecided membership. They also obtained low-dispersion spectra for 41 faint candidate stars, of which 25 showed definite H$\alpha$ emission while 5 showed possible weak H$\alpha$ emission. P93 include these 30 stars as probable members of IC~4665.

Multiple investigations into chromospheric and coronal activity have been conducted in IC~4665. H$\alpha$ emission, as an indicator of chromospheric activity, has also been observed in 14 F-, G-, and K-spectral type stars \citep{Martin1997}, as well as for 39 lower-mass, M-spectral type stars (P93). More recently, \citet{Jeffries2009} measured H$\alpha$ equivalent widths for 56 IC~4665 stars ranging from early-F to mid-M spectral type. In addition to chromospheric studies, IC~4665 also has been observed with the ROSAT X-ray telescope, in order to detect coronal X-ray emission. \citet{Giampapa1998} reported the results of a 76.9 ks exposure of the cluster with the HRI instrument on ROSAT, from which they found 28 X-ray detections matching optical counterparts in the P93 catalog. Six (6) of these 28 X-ray sources were shown to be cluster nonmembers by \citet{Giampapa1998} based on spectral-type classifications.

IC~4665 is currently one of just five open clusters with a detected lithium depletion boundary (hereafter LDB). \citet{Manzi2008} measured an age of $28\pm5$ Myr for the cluster based on the absolute magnitude of stars found at the LDB. Moreover, the global lithium abundance distribution in IC~4665 stars was first studied by \citet{Martin1997}, where they measure the Li abundance of 14 early-G through early-M spectral type stars. Interestingly, they found that the Li distribution of IC~4665 is similar to the older Pleiades open cluster ($\sim$125 Myr), although they had a limited sample size of measured Li abundances. More recently, \citet{Jeffries2009} determined the Li abundance of 20 late-F to early-M spectral type stars, and confirmed that the Li distribution of IC~4665 appears more like the Pleiades than other younger open clusters. 

The stellar rotation distribution of IC~4665 is commonly used to empirically constrain theories of early angular momentum evolution \citep[e.g.][]{Barnes2003a,Barnes2007}. Using rotation period measurements for nine solar-type stars (F9 to K0 spectral type), \citet{Allain1996} find that there is a dearth of slow rotators (periods $\geq$ 4 days) in IC~4665, indicative of a young cluster where the stars have not had time to spin down. However, \citet{Jeffries2009} measure projected rotation velocity (V$\sin{i}$) for IC~4665 cluster members, and find that there is a lack of fast rotators (V$\sin{i} > 20$ km s$^{-1}$) with 1 $<(V-I_{c})<$ 2 ($\sim$ K-type stars). Comparing this V$\sin{i}$ distribution to the older IC~2391, IC~2602, and Pleiades, they find that the K-type stars in IC~4665 appear to rotate more slowly than one would expect at $\sim$30 Myr. Finally, a recent survey reported rotation periods for 20 IC~4665 members with masses $<$ 0.5$M_{\sun}$ \citep{Scholz2009}. They show that all of these low-mass stars are rotating rapidly (period $\leq$ 1 day), a finding similar to studies of other PMS clusters \citep[e.g.][]{Irwin2008}.   

\subsection{New SMARTS BVI$_{c}$ Observations}\label{sec.newphot}

\begin{figure}[h] 
  \centering
  \includegraphics[scale=0.5,angle=0]{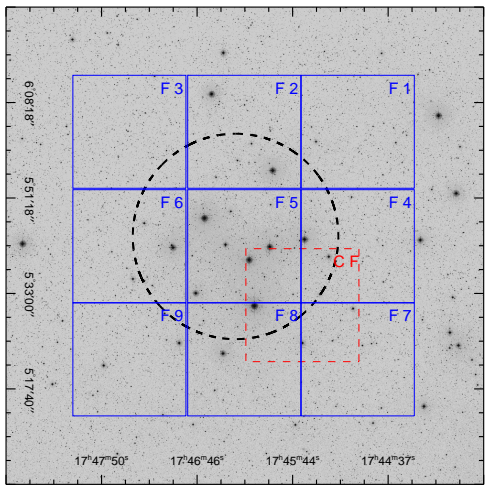} 
  \caption{
    \label{fig.dssimage}  
  Digital Sky Survey image of the areal region covered by our SMARTS IC~4665 survey. 
  Each square represents an observed field (coordinates given in Table~\ref{tab.pointings}). 
  Also plotted is the overlapping control field (CF) identified by the red dashed box, and 
  the cluster's core radius (black dashed-circle) as derived by \citet{Kharchenko2005}.
  } 
\end{figure}

We observed IC~4665 using the 1.0m SMARTS telescope (the old YALO telescope) situated at the Cerro Tololo Interamerican Observatory (CTIO), Chile, during the nights of 15, 16, 17 September, 2005. The data were acquired with the quad-amplifier Y4KCam CCD camera, equipped with Johnson-Cousins BVI$_{c}$ filters, with its 15 $\mu$m pixels and 0.289 arcsec/pixel plate scale results in an on-sky field of view per CCD field of $19\farcm3 \times 19\farcm3$. Nine fields were observed covering a uniform grid of a $3 \times 3$ mosaic of CCD fields covering approximately 1 square degree, centered on R.A.(2000):$17^{h} 46^{m} 13.5^{s}$ and Dec(2000):$+05\arcdeg 41\arcmin 52\arcsec$. We employed exposure times of 100, 50, and 25 seconds for the B-, V-, and I$_{c}$-band observations, respectively. The central coordinates of each IC~4665 field for which BVI$_{c}$ photometry was obtained are shown in Fig.~\ref{fig.dssimage} and summarized in Table~\ref{tab.pointings}. 

 \input{tab1.tex}

All science and standard star images were processed for overscan region subtraction, master-bias subtraction and flat fielding, using twilight sky images, using standard procedures in the IRAF\footnote{IRAF, in our case through http://iraf.net, is distributed by the National Optical Astronomy Observatories, which are operated by the Association of Universities for Research in Astronomy, Inc., under cooperative agreement with the National Science Foundation.} suite of data reduction algorithms. Substantial use was also made of two batch processing IRAF scripts, written by Phil Massey of Lowell Observatory, whose protocol specifically handles the FITS headers and quad-amplifier readout of the CCD camera. The relatively low space density of the cluster, and its lack of substantial field crowding, allows us to employ aperture photometry for both IC~4665 science and standard star images. Source searching and aperture photometry, with an aperture radius of 13-pixels, were achieved using the DAOPHOT package in IRAF \citep{Stetson1987,Stetson1992}. 

Equatorial BVI$_{c}$ standard stars cataloged in \citet{Landolt1992,Landolt2009}, with a broad dynamic range of photometric magnitude and color, and located at wide distribution of airmass, were observed nightly in order to correct for atmospheric extinction and to transform extinction corrected, instrumental magnitudes onto the standard system. Modified forms of Bouguer's law, see Eqs.\ref{eqn:phot}~(a-d), were used to define the relationship between standard and instrumental magnitudes for the standard stars. 

\begin{mathletters}
\label{eqn:phot}
\begin{eqnarray}
V = v + \varepsilon (B-V) + \xi_{v} - \kappa_{v}~X  \\
V = v + \varepsilon (V-I_{c}) + \xi_{v} - \kappa_{v}~X  \\
(B-V) =  \mu  (b-v) + \xi_{bv} - \mu~\kappa_{bv}~X \\ 
(V-I_{c}) = \psi (v-i_{c}) + \xi_{vi} - \psi~\kappa_{vi}~X 
\end{eqnarray}
\end{mathletters}

\hspace*{4mm} where
\begin{itemize}
\item
V, $(B-V)$ and $(V-I_{c})$ are magnitudes/indices on the BVI$_{c}$ standard system 
\item
v, $(b-v)$, and $(v-i_{c})$ are measured, instrumental magnitudes/indices
\item
$\kappa_{v}, \kappa_{bv}$ and $\kappa_{vi}$ are filter dependent extinction coefficients 
\item
$\varepsilon$, $\mu$ and $\psi$ are color transformation coefficients 
\item
$\xi_{v}$, $\xi_{bv}$ and $\xi_{vi}$ are zero-point coefficients.
\item 
X = airmass of target -- $\simeq$ sec {\it z} - where {\em z} is the zenith distance
\end{itemize}

\input{tab2.tex}

Unknown extinction coefficients, color transformation coefficients and zero points were determined by solving the self-similar series of linear simultaneous algebraic equations detailed in Eqs.\ref{eqn:phot}~(a-d). Solutions were derived using a least-squares fit algorithm\footnote{The algorithm, F04AMF, is distributed by the Numerical Algorithms Group (NAG)} which processes the numeric parameters via an iterative refinement method. Comparing calculated magnitudes and colors of Landolt standard stars to their published values allows an iterative rejection process to take place, thereby eliminating seriously discrepant standard star measurements -- of which there were few. Solutions to the color equations for each observing night are listed in Table~\ref{tab.trans}, whereby these solutions resulted in differences between calculated and published magnitudes of Landolt standard stars that were never greater than $2.2\%$ ({\em i.e.,} 0.022 magnitudes). Target star photometry was readily determined by substituting the derived extinction, transformation and zero-point coefficients along with measured instrumental magnitudes into Eqs.\ref{eqn:phot}~(a-d). 

\begin{figure}[h] 
  \centering
  \includegraphics[scale=0.4,angle=90]{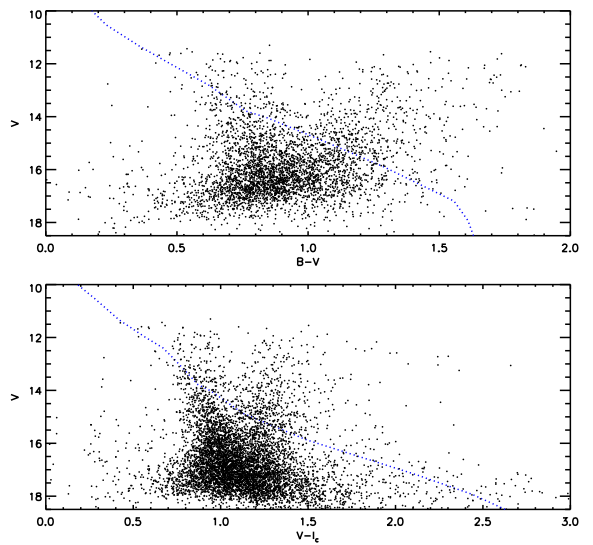} 
  \caption{
    \label{fig.cmds}  
    Color-magnitude diagrams for our full photometric catalog are 
    plotted (black dots). Also provided are \citet{DAntona1997} 
    isochrones  (blue dotted line) for the cluster's lithium depletion 
    boundary age \citep[27.8 Myr; ][]{Manzi2008} and HIPPARCOS 
    distance \citep[355 parsecs; ][]{vanLeeuwen2009}.
  }
\end{figure}

The full BVI$_{c}$ catalog for all IC~4665 fields is provided in Table~\ref{tab.full}, including their associated 2MASS J, H, and K$_{s}$ photometric and astrometric data, correlated with point sources in the online 2MASS database \citep{Skrutskie2006}. Of the 7359 stars in our master BVI$_{C}$ catalog, we find 452 stellar sources without matches within 3.0 arcseconds of 2MASS sources. For these stars, we provide J2000 astrometry, precise to $\pm 1$ arcsecond, using a 6th-order polynomial fit to X,Y CCD coordinates and accurate and precise coordinates for reference stars in the SuperCOSMOS catalog \citep{Hambly2001}. A visual presentation of the BVI$_{c}$ photometric catalog is shown in Fig.~\ref{fig.cmds}, in which we plot V versus $B-V$ and $V-I_{c}$ CMDs in concert with \citet{DAntona1997} fiducial isochrones.

\input{tab3a.tex}
\input{tab3b.tex}

\begin{figure} 
  \centering
  \includegraphics[scale=0.4,angle=90]{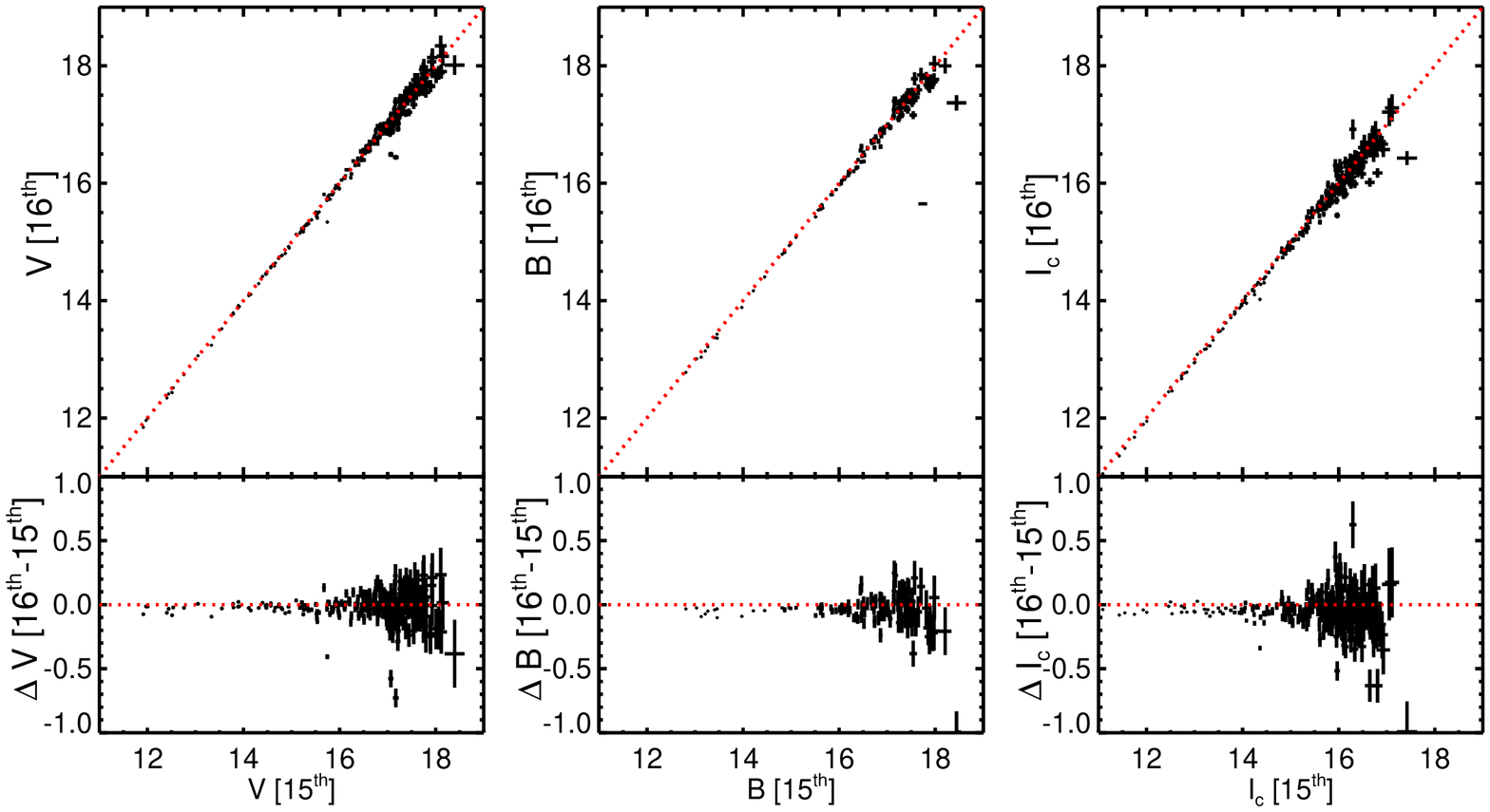} 
  \includegraphics[scale=0.4,angle=90]{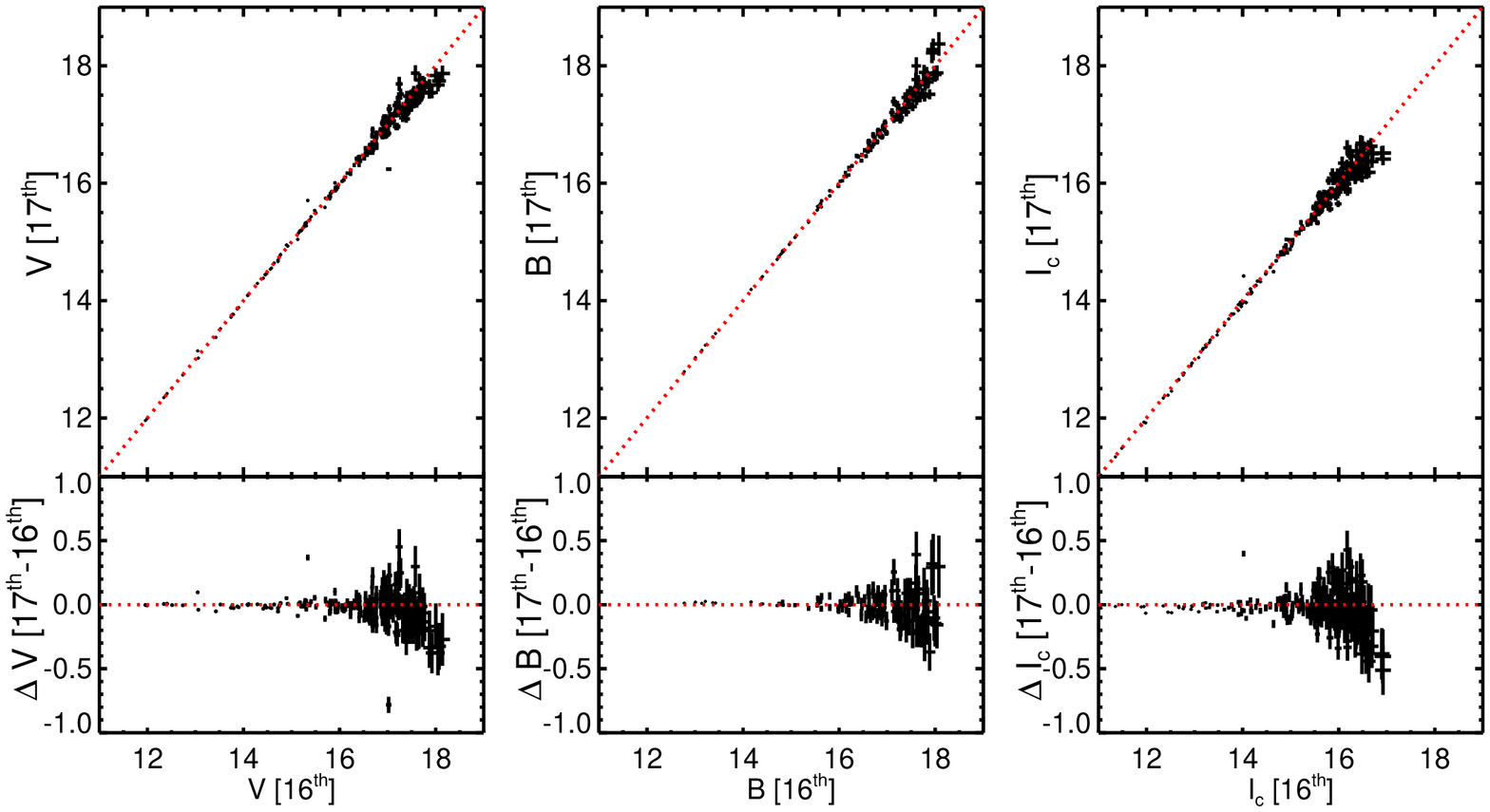} 
  \caption{
    \label{fig.internal}  
    Comparison of CTIO BVI$_{c}$ photometry for control fields in IC 4665, based on 
    observing nights 15, 16, and 17 Sept. 2005. Upper panels show direct comparison, while 
    their residuals are plotted in each lower one. Data are plotted with their associated 
    photometric error. Red dotted lines are loci of equality and not fits to the data.
  }
\end{figure}

In order to investigate the stability of our photometric system, we also observed a control field (CF) in IC~4665 during the nights of the 15$^{th}$ \& 17$^{th}$ of September 2005. Fig.~\ref{fig.internal} shows our V-, B-, and I$_{c}$-band magnitude comparisons for stars in common to the control fields observed on the two different nights. A simple statistical analysis of these data is presented in the Table~\ref{tab.internal}. From these plots and table, we can infer that the V- and I$_{c}$-band zero points are at least consistent at the level of $\sim3-4\%$ for objects with V and I$_{c} < $16 magnitudes. For these same objects, the B-band observations do however show an offset at the $\sim5-6\%$ level. Naturally, for the dimmer objects (V, B, I$_{c}>$16 magnitudes), with decreasing S/N levels, diminished precisions of order $\sim9-10\%$ are seen.

\input{tab4.tex}

\begin{figure}
  \centering
  \includegraphics[scale=0.65,angle=0]{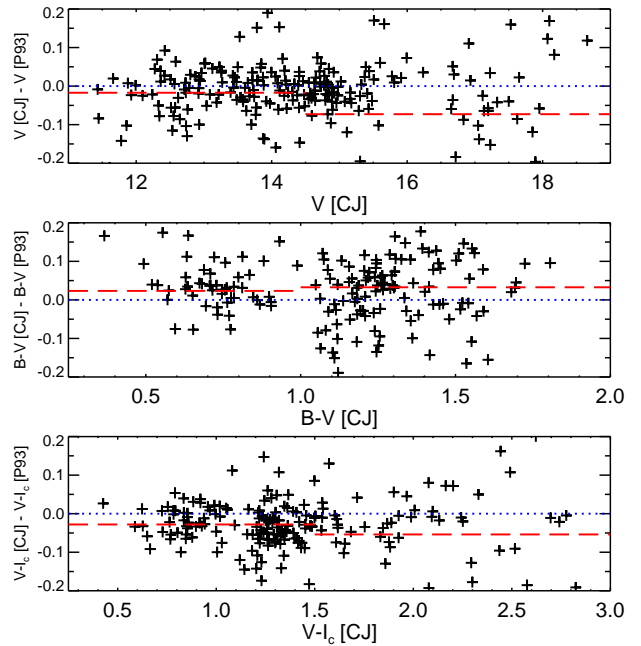} 
  \caption{
    \label{fig.P93comp}  
    $V$, $B-V$, and $V-I_{c}$ residual comparison plots for the 
    P93 catalog and our optical catalog (CJ). Our translation from 
    P93 $V-I_{k}$ to $V-I_{c}$ is described in Section \ref{sec.comp}. 
    The blue dotted lines indicate equality between the systems, whereas 
    the red dashed lines are calculated means for V $<$ 14.5 \& 
    $\geq$ 14.5; $B-V < 1$ \& $\geq$ 1; $V-I_{c}$ $<$ 1.5 \& $\geq$ 1.5, respectively. 
  } 
\end{figure}

\subsection{Comparison to Prosser (1993) Photometry}\label{sec.comp}

Existing studies reporting on the physical characteristics of IC~4665 solar-type stars are under-pinned on the P93 photometry study, and it is therefore imperative that we determine the reliability of their photometric system as compared to ours. A detailed examination of the standardization procedure of P93 shows that, although nightly standard stars were used to derive extinction coefficients and color-transformation equations, the zero-points of their BVI$_{k}$ photometric system are tied to a solitary blue, high-mass B9e star. This star, HD~161261, has subsequently been found to have V-band variability on the order of 0.1 mag \citep{Kazarovets1997}, which maybe due to a binary component \citep{Pourbaix2004}, as well as an infrared excess typical of rapidly rotating Be stars \citep{Yudin2001}. Not surprisingly, we observe systematic offsets between the P93 photometry and our own that are likely a result of this extrapolation of standardized photometry derived from such a single, blue, possibly variable and binary standard star.

For stars in common to both catalogs, we plot in Fig.~\ref{fig.P93comp} comparisons between the P93 study and our own $V$, $B-V$, and $V-I_{c}$ photometric datasets. We translated the P93 Kron $V-I$ colors to the Cousins system using the formalism given in \citet{Bessell1987}. We see a significant amount of scatter in the P93 photometry ($\sim$0.1 mag in V; $\sim$0.2 in $B-V$/$V-I_{c}$) compared to our optical catalog. In addition, there appears to be a systematic offset of $\sim$0.05 mag in V, $B-V$, and $V-I_{c}$ in the faintest/reddest P93 objects. Such a magnitude of photometric scatter and systematic offset may cause significant problems when trying to calibrate both color and magnitude relationships for stars in IC~4665; therefore, for the purpose of this manuscript, we will preferentially employ our, new BVI$_{c}$ photometry in all subsequent analysis.

\section{The Age of IC~4665}\label{sec.age}

Determining a precise and accurate age for IC~4665 is fundamental to its use as a constraint, or boundary condition, for theoretical models describing early stellar evolution. Its age however, continues to be a matter of uncertainty. Ages for IC~4665 found in the literature range from greater than 25 to less than 100 Myr, although one of the age determinations is derived using the lithium depletion boundary method \citep{Manzi2008}, which so far has only been applied to a small handful of open clusters. With such a large range of possible ages cited for IC~4665, its applicability for constraining theoretical stellar evolution models remains suspect. Here, we use our new photometric data for cluster members of IC 4665 to take a new look at its discrepant age determinations to verify whether there exists a true, observational-dependent age spread in IC~4665, or whether this spread in ages is due to uncertainty in the extant studies of the clusters.

\subsection{Pre-Main-Sequence Age}\label{sec.lowmass}

In the P93 study, the authors reported that they were unable to determine the age of IC~4665 from the location of the low-mass pre-main-sequence, due to the lack of a defined low-mass sequence in their CMD. This may have been the result of systematic scatter in their photometry and/or their cluster membership determination being based on unreliable proper motion data. More recently, the infrared study of IC~4665 by \citet{deWit2006} suggests that the cluster has an age of 50 to 100 Myr based on the location of its low-mass cluster sequence. However, \citeauthor{deWit2006} note that an accurate age for the cluster is difficult to measure due to the large amount of field star contamination present in their survey. 

\begin{figure*}
  \centering
  \includegraphics[scale=0.8,angle=0]{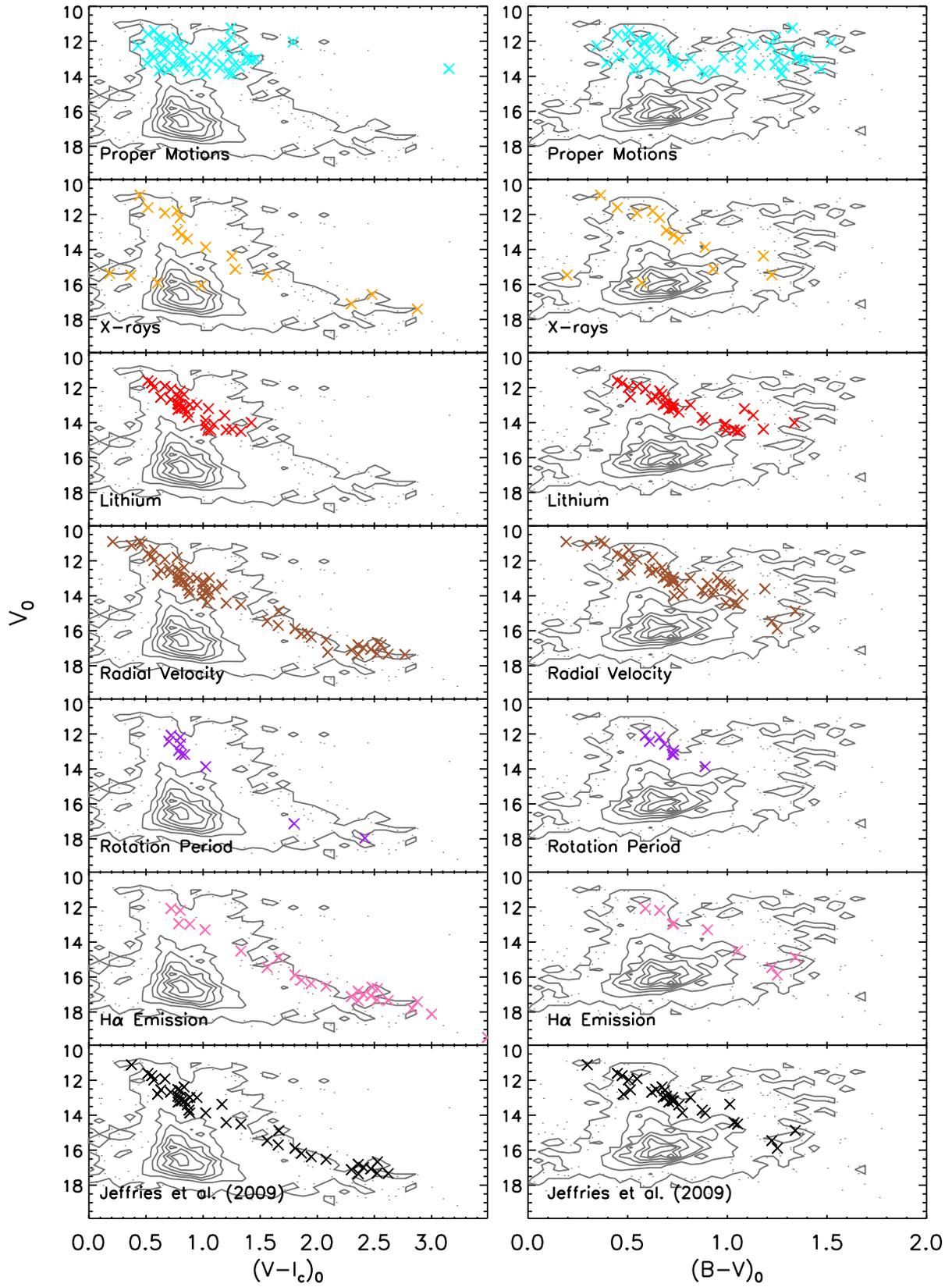} 
  \caption{
    \label{fig.multicmd}  
    Intrinsic $V-I_{c}$ and $B-V$ color-magnitude diagrams for IC~4665 are 
    presented (object density given by gray contours) with cluster candidate 
    members identified (crosses) using a secondary membership criterion, 
    labeled in the lower-left corner of each panel (see Section 
    \ref{sec.lowmass} for source references). We use a cluster reddening 
    of $E(B-V) = 0.174$ (WEBDA), an $E(V-I) = 1.25 \times E(B-V)$, and an 
    A$_{v} = 3.12\times E(B-V)$ relation, in order to derive intrinsic colors 
    and magnitudes. Colors in each panel correspond to the associated data 
    points in Figure \ref{fig.bestfit}. 
    }
\end{figure*}

A quick examination of the full IC~4665 CMD reveals that we too confirm that an accurate fit of the cluster sequence is particularly difficult to achieve due to field star contamination in our color-magnitude diagram (see Fig.~\ref{fig.cmds}). While IC~4665 lies relatively far from the Galactic plane ({\it b} $=+17\degr$) the level of contamination along the CMD suggests that there is a high stellar density in the cluster's line of sight through the Galaxy. In order to account for the field star contamination in the IC~4665 CMDs, we further constrain our analysis to those stars having ancillary membership indicators from a disparate range of extant photometric and spectroscopic observing campaigns. In the first instance, we positionally matched our photometric catalog with data from the following sources: three catalogs of stars with common kinematic properties (proper motions --- P93; radial velocities --- \citealt{Prosser1994,Jeffries2009}), stars with X-ray emission \citealt{Giampapa1998}, two catalogs of stars with significant lithium absorption \citep{Martin1997,Jeffries2009}, three catalogs of stars exhibiting strong H$\alpha$ emission (EW$_{H\alpha} > 0~\AA$ --- P93; \citealt{Martin1997,Jeffries2009}), and finally, two catalogs of stars with short rotation periods in the region of IC~4665 \citep{Allain1996,Scholz2009}. X-ray and H$\alpha$ emission (coronal and chromospheric), the presence of lithium in stellar photospheres, and rapid rotation are all indicators of youth for low-mass stars, and therefore we can exploit such surveys in order to segregate young cluster members from the typically much older field star populations. In Fig.~\ref{fig.multicmd}, we plot the CMDs with stars identified from the various membership catalogs. We will subsequently exclude proper-motion-selected members from our cluster catalog due to the lack of a clear (pre-) main-sequence in its CMD (see Fig.~\ref{fig.multicmd}; top panels), {\em viz}, proper motion must be an extremely poor membership criterion for this cluster. This can be explained by the similarity of the cluster's systemic proper motion to that of the general field population, justifying the exclusion of the cluster's proper motion catalog from all subsequent analysis of IC~4665 \citep[see][]{Jeffries2009}. 

We recognize that the membership catalogs cited above contain varying levels of field contamination, as well as biases based on their color and/or magnitude ranges. For instance, although our catalog of radial velocity members contains the largest number of stars spread over the largest range of colors, it also has a high probability of field star contamination due to the similarity of the kinematic distribution of the cluster compared to that of the field \citep{Jeffries2009}. On the other hand, the catalog of stars showing X-ray emission has a much lower probability of contamination due to the ubiquitous nature of high magnetic activity in young stars compared to old field stars. However, the relatively low number of stars identified can be attributed to the fairly narrow field-of-view of the ROSAT X-ray satellite compared to the on-sky distribution of apparent cluster members. In an attempt to account for the relative power of using multiple membership criteria, \citet{Jeffries2009} produced a catalog of highly probable members of IC~4665 using a combination of kinematic motions, lithium content, optical and 2MASS photometry, and spectral-index indicators as filters to remove foreground and background field star interlopers in the cluster catalog. We include this list of high-fidelity cluster members in our analysis, and note that it is most current, up-to-date compilation of {\em bona fide} IC~4665 members.

\begin{figure} 
  \centering
  \includegraphics[scale=0.5,angle=0]{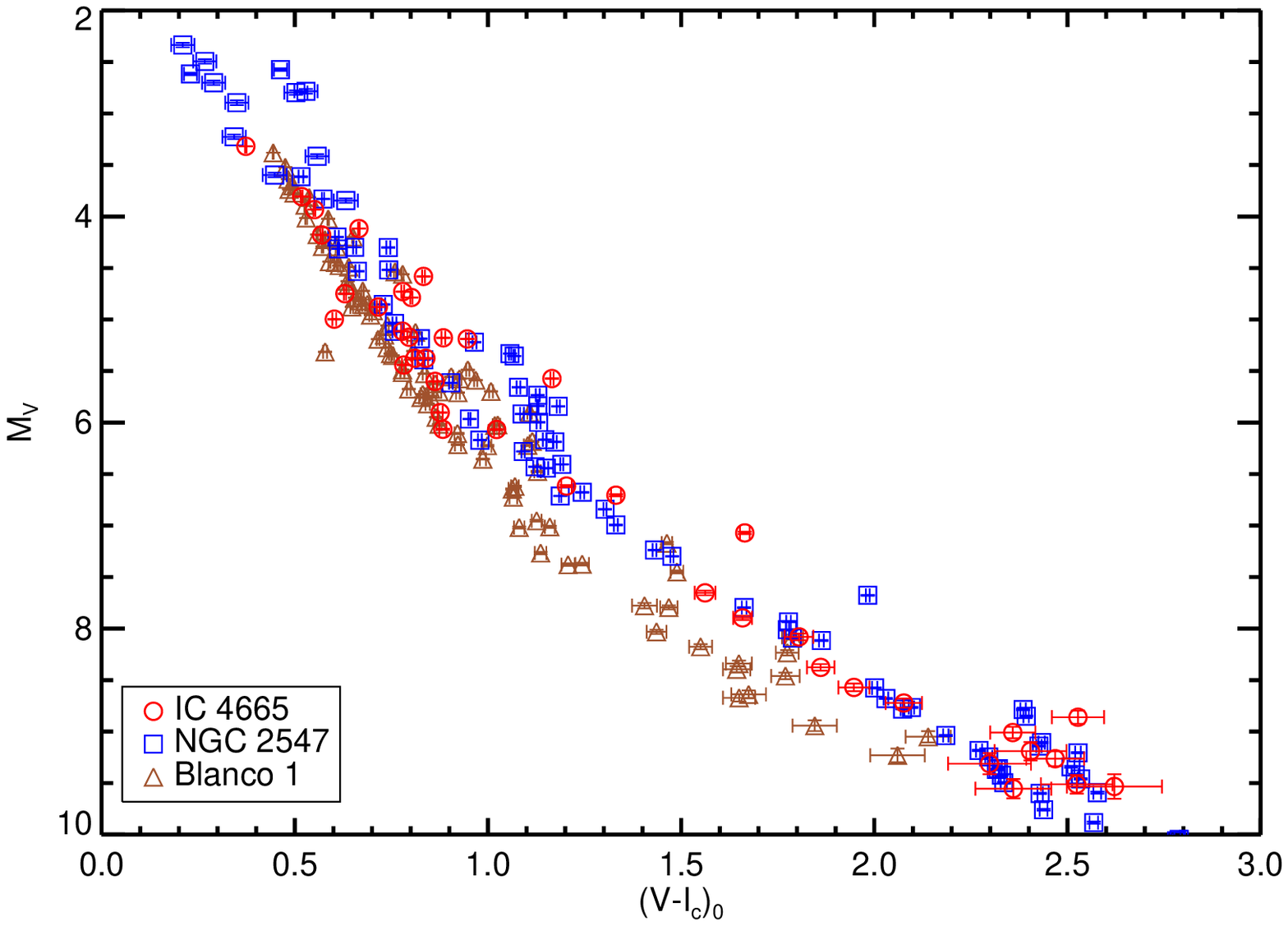} 
  \caption{
    \label{fig.isocomp}  
    Absolute $V$ versus intrinsic $V-I_{c}$ magnitudes are plotted for 
    the open clusters IC~4665, NGC~2547 (35 Myr), and Blanco~1 (100 Myr). 
    Intrinsic colors and magnitudes were calculated using cluster's 
    $E(B-V)$ data (WEBDA), an $A_{v} = 3.12 \times E(B-V)$ and an 
    $E(V-I_{c}) = 1.25 \times E(B-V)$ relation. The apparent-to-absolute 
    magnitude conversion was achieved using the appropriate HIPPARCOS 
    distances reported in \citet{vanLeeuwen2009}. See 
    Section~\ref{sec.lowmass} for data sources.
    }
\end{figure} 

In an effort to find an initial estimate for the age of IC~4665 using its low-mass population, we plot in Fig.~\ref{fig.isocomp} the $V$ versus $V-I_{c}$ color-magnitude diagram for confirmed members of three open clusters; for IC~4665, the membership catalog from \citet{Jeffries2009} is used, whereas the 35 Myr old NGC~2547 cluster, we employ the photometric membership catalog from \citet{Naylor2002}. Finally for the $\sim$100 Myr old Blanco~1 cluster, James et al. ({\em in prep}) provide us with its photometric catalog of kinematic members. While some photometric scatter is apparent, probably due to binarity and a small number of cluster non-members, these CMDs evidently show that the lower mass stars in IC~4665, $(V-I_{c})_{o}>1.00$, lie systematically above the Blanco~1 main sequence. In fact, in this color range IC~4665 stars have colors and magnitudes closely matching the corresponding stars in NGC~2547, suggesting that at least photometrically, IC~4665 is of a comparable age to NGC~2547, {\em i.e.,} $\sim 35$ Myr. Although this age estimate is insightful, we now establish a more robust determination of the cluster's age using the so-called $\tau^{2}$ isochrone fitting technique \citep{Naylor2006}.

\subsubsection{$\tau^{2}$ Isochrone Modeling}\label{sec.tau2}

The classic approach in determining fundamental properties of a young open cluster, such as age and distance, is finding a best-fitting CMD isochrone for the apparent cluster sequence. Typically, this is performed using a ``by-eye'' comparison of the model isochrone to the observational data, which can be burdened by uncertainties due to field star contamination and/or photometric error, as well as an apparent spread in the cluster sequence due to binarity. In this manuscript, we describe a new, statistical approach in modeling the lower-main-sequence of IC~4665. 

The Tau-squared ($\tau^{2}$) maximum likelihood statistic has recently been developed for modeling CMDs with two-dimensional theoretical isochrones \citep{Naylor2006,Naylor2009}. Examples of its application can be found in \citet{Jeffries2007b,Joshi2008,Mayne2008,Jeffries2009b}. In short, $\tau^{2}$ is a generalized $\chi^{2}$ statistic that takes into account correlated uncertainties in two dimensions. Thus, we can use this statistical tool to effectively model a two-dimensional distribution, in our case an age- and distance-dependent CMD isochrone, to the $V/B-V$ and $V/V-I_{c}$ CMDs by minimizing $\tau^{2}$ to find a best-fitting isochrone. 

Following closely the technique described in \citet{Naylor2006,Naylor2009}, we produce a grid of stellar model isochrones spanning a range of ages and distances. We use the low-mass, solar-metallicity stellar evolution models from \citet{DAntona1997}, \citet{Baraffe1998}, and \citet{Siess2000}, and employ a {\em Pleiades tuning} method to translate model luminosities and effective temperatures ($T_{eff}$) into magnitudes and colors \citep[for a full description of the {\em Pleiades tuning} method, see][]{Jeffries2001,Naylor2002}. The $\tau^{2}$ software\footnote{The $\tau^{2}$ software is publicly distributed by T.~Naylor at http://www.astro.ex.ac.uk/people/timn/tau-squared.} uses Monte Carlo methods to produce a probability, based on $\tau^{2}$, of finding a star drawn from the full photometric catalog (including photometric binaries) represented by each age-distance pixel on the grid. These probabilities are calculated for all stars in the CMD, and summed to derive the overall $\tau^{2}$ for the given age and distance isochrone. By repeating this process for a range of ages and distances, a $\tau^{2}$ grid is produced, and the minimum value is the best-fit isochrone for the modeled CMD. 

In order to have our best-fit isochrones not be influenced by obvious outliers, we added an additional {\em soft-clipping} rejection step \citep{Naylor2006}, where by points having colors/magnitudes lying several $\sigma$ away from the range of model isochrones are {\em ab initio} removed from the modeling process. In our modeling of IC~4665, we assume an extinction relation of $A_{v} = 3.12 \times E(B-V)$, a reddening of $E(B-V)=0.174$ (WEBDA Open Cluster Database), an $E(V-I_{c}) = 1.25 \times E(B-V)$ relation and a binary fraction of 0.5 \citep{Crampton1976}.

We have assumed in our analysis that differential reddening is negligible for IC~4665. This is based on the results of \citet{Crawford1972} study finding the reddening in $B-V$ varied by only $\sim$0.02 magnitudes for 17 high-mass (B spectral-type) stars across the cluster. This amount of variation in $B-V$ color excess is at the level of our photometric internal precision ($\Delta (B-V) \sim 0.017-0.022$ mag) as determined by our nightly comparison analysis (see Section \ref{sec.newphot} and Table \ref{tab.internal}).

Once a best-fit isochrone for each of our IC~4665 CMDs is found, we derive the probability of the isochrone model being a good fit based on their minimum $\tau^{2}$ values. We include an additional magnitude-independent systematic uncertainty to the colors and magnitudes in our CMDs to increase their probabilities to an acceptable value (i.e.,$\gtrsim30\%$). This is analogous to the traditional method of increasing error bars to arrive at a $\chi^{2}_{\nu} \sim 1$, and ensures that the uncertainties on our ages and distances are appropriate relative to the scatter in the IC~4665 CMDs. Typically, we found that an additional $\sim 0.01 - 0.03$ increase is required for our V, B, and I$_{c}$ magnitudes to achieve sufficiently high best-fit probabilities. This magnitude of systematic uncertainty can be easily explained by systematics in the standardization procedure \citep{Naylor2002} or natural variability in such young, solar-type stars \citep[{\em e.g.,} spotted stars in IC~4665 have reported optical variability amplitudes of $\sim 0.03 - 0.1$ mag,][]{Allain1996}.

\begin{figure}
  \centering
  \includegraphics[scale=0.5,angle=0]{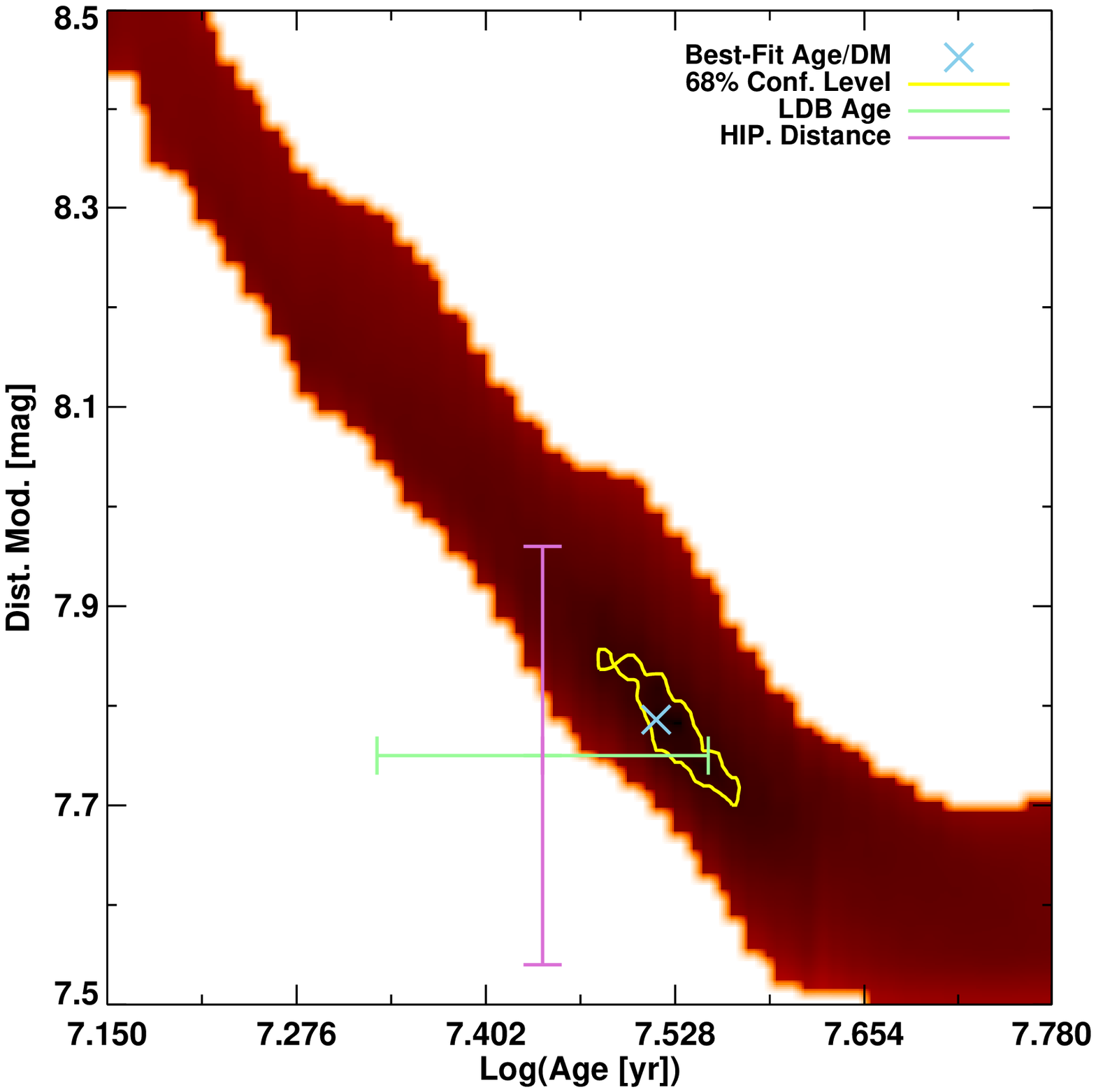} 
  \caption{
    \label{fig.taugrid}  
    $\tau^{2}$ space resulting from modeling the \citeauthor{Jeffries2009} membership catalog 
    with \citeauthor{DAntona1997} isochrones. The color-scale is proportional to the $\tau^{2}$ 
    value for the given ages and DM. The yellow contour is 68\% confidence level and the blue 
    cross indicates the minimum $\tau^{2}$ value, i.e., the best-fit age and DM. For comparison 
    purposes, the lithium depletion boundary age \citep{Manzi2008} and the HIPPARCOS distance 
    \citep{vanLeeuwen2009} are also plotted with their respective uncertainties. 
    }
\end{figure}

\begin{figure}
  \centering
  \includegraphics[scale=0.5,angle=0]{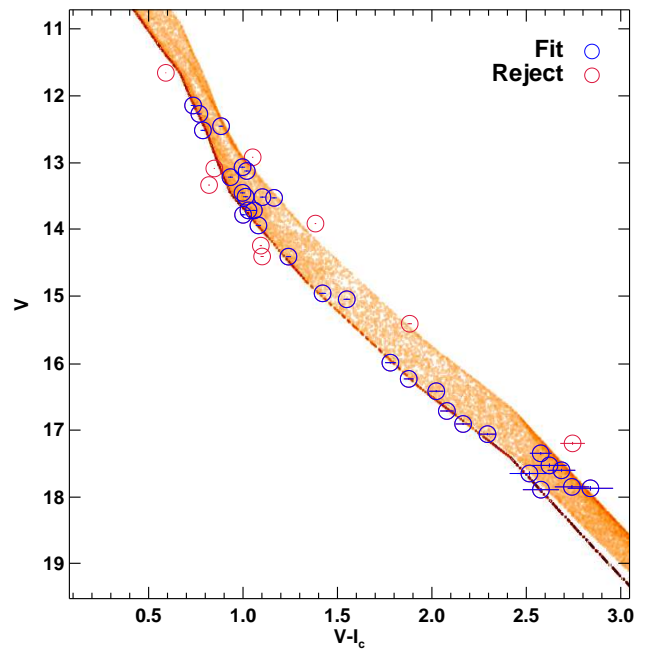} 
  \caption{
    \label{fig.examplecmd}  
    Color-magnitude diagram for the objects in the \citeauthor{Jeffries2009} membership catalog. 
    Plotted is the best-fit \citeauthor{DAntona1997} isochrone from our $\tau^{2}$ analysis, with 
    the color-scale indicating the simulated relative number density of stars along the cluster 
    sequence. The red circles indicate objects that were rejected by our ``soft-clipping'' scheme, 
    and the blue circles are the data used in the isochrone modeling.
    }
\end{figure}

As an example of the basic procedure behind $\tau^{2}$ isochrone modeling, we plot in Fig.~\ref{fig.taugrid} the $\tau^{2}$ grid derived from $V-I_{c}$ CMD isochrone modeling of the \citet{Jeffries2009} membership catalog using the PMS models from \citeauthor{DAntona1997}. The displayed color-scale indicates the total $\tau^{2}$ values for the indicated distance modulus (hereafter DM) and age grid points. The minimum value in this grid defines the best-fit age and DM for this membership catalog. This grid of $\tau^{2}$ values can also be thought of as a grid of probabilities (see above); therefore, $\tau^{2}$ values can be found for different confidence levels, e.g., displayed in Fig.~\ref{fig.taugrid} is the 68\% confidence contour. The ranges of ages and distances falling within this contour defines the uncertainty in our best-fit measurements. We present in Fig.~\ref{fig.examplecmd} the best-fit isochrone from this $\tau^{2}$ grid analysis. 

\subsubsection{Pre-Main-Sequence Isochrone Results}\label{sec.pmsresults}

\begin{figure*} 
  \centering
  \includegraphics[scale=0.7,angle=0]{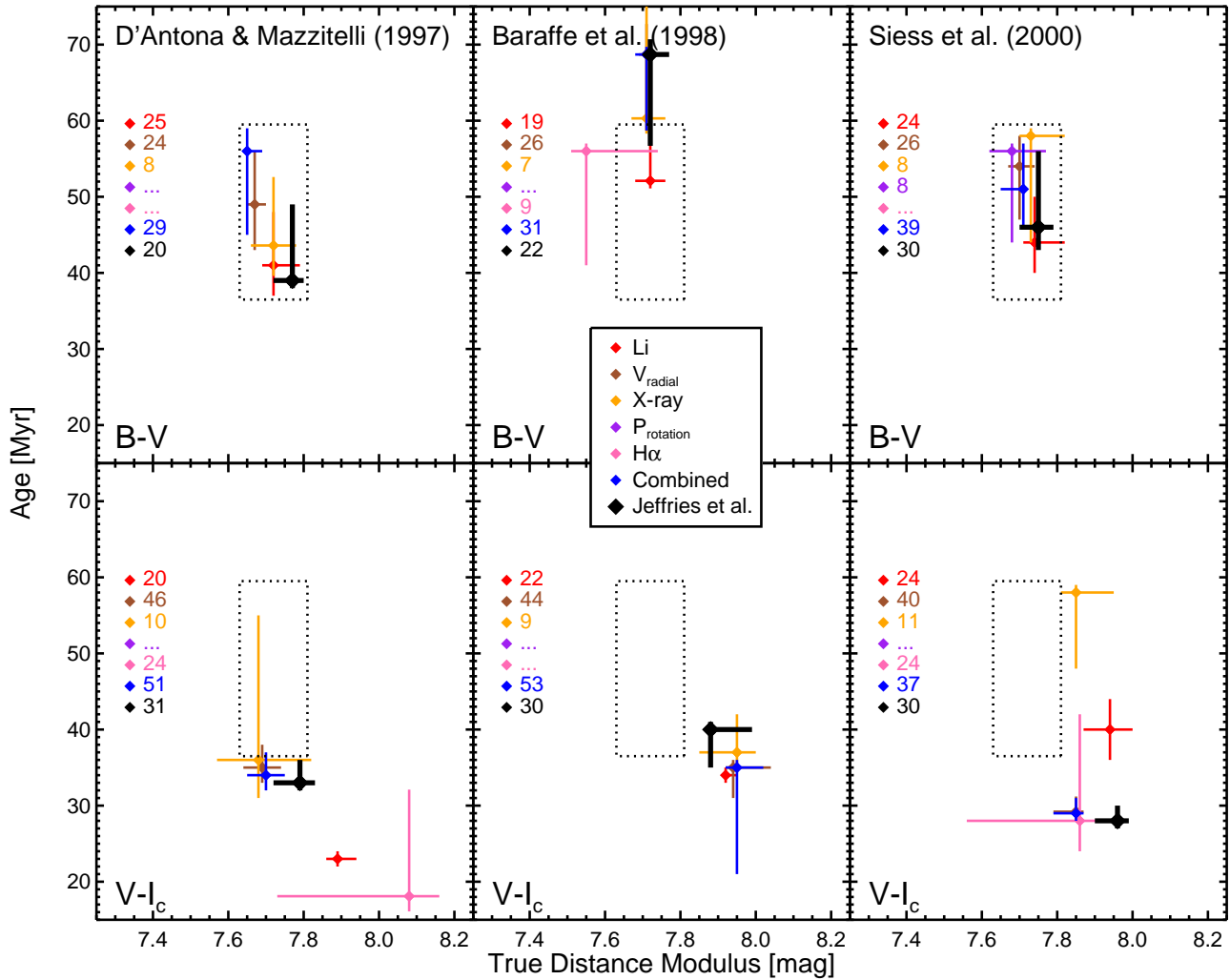} 
  \caption{
    \label{fig.bestfit}  
    Best-fit ages and intrinsic distance moduli resulting from our $\tau^{2}$ 
    isochrone modeling of the low-mass members of IC~4665. Upper panels 
    represent results for $(B-V)$ CMDs, while the lower panels represent 
    the $(V-I_{c})$ ones. Each data point's color is described by the legend in 
    the center of the plot (see Section~\ref{sec.lowmass} for source 
    references). Along the left of each plot, the final number of data points 
    are provided that were used in the CMD modeling after having 
    applied a soft clipping process. Data sets for which we were unable 
    to find a best-fit age and DM, due to the $\tau^{2}$ model isochrone 
    not converging, are identified by '...' instead of the number of data 
    points used given on the left. For comparison purposes, we plot a black 
    dashed-line box in each box, which represents the approximate range of 
    ages and DM that result from the modeling of the $B-V$ CMD using the 
    \citeauthor{DAntona1997} isochrones (upper left panel).
  }
\end{figure*}

In Fig.~\ref{fig.bestfit} we show the best-fit ages and distances, derived from our Tau-squared fitting procedure, for each of our seven IC~4665 membership catalogs (5 individual catalogs described previously, the catalog from \citet{Jeffries2009}, and a combined catalog that includes any object with some type of membership criterion), and three stellar evolution models representing two different colors ($B-V$ and $V-I_{c}$). We note that we were unable to find a best-fit age and DM solution for every membership catalog using each model and/or color, with the rotation period and H$\alpha$ catalogs being particularly problematic. The failure of these model convergences is primarily due to the specifics of each membership catalog ({\em e.g.,} large amounts of field-star contamination, not having a wide color-magnitude range in data points, {\em etc}.). 

\input{tab5.tex}

In the following analysis we choose to use the membership catalog from \citet{Jeffries2009} as the exemplar best-fit age and DM for the different models and colors, as it has the lowest probability of contamination by field stars (see Section \ref{sec.lowmass}). Nevertheless, modeling multiple membership catalogs allows us to identify the possible range of ages and DM one should expect when using specific membership criteria to identify cluster members. In Table \ref{tab.isoresult}, we list the median of the ages and distances, derived for all membership criterion catalogs, together with the observed range of values from our best-fit isochrone for each color and stellar model. We also include the best-fit solution for the \citeauthor{Jeffries2009} membership list. 

For the most part, the best-fit ages and DM found for all three PMS models are typically within their individual errors of each other, suggesting that statistically they are all three are in agreement of IC~4665's age and distance. However, we do find that some points lie beyond 1 $\sigma$ from the other derived ages and DM ({\em e.g.}, the best-fit parameters found for the $V-I_{c}$ Li catalog using the \citeauthor{DAntona1997} models), and are most likely the result of a poor fit due to limited or discrepant data. This is evidence for the strength of modeling multiple membership catalogs, that we are able to clearly adjudge whether an individual catalog's measured age and DM is discrepant due to a poor isochrone fit or otherwise.

Here, we would like to highlight some results of our PMS isochrone modeling of IC~4665. First, if we focus solely on the $B-V$ isochrones, we find good agreement in the measured DM for all three models, whereas only the ages for the \citeauthor{DAntona1997} and \citeauthor{Siess2000} models are in agreement with the best-fit values for \citeauthor{Baraffe1998} model isochrones yielding systematically older ages (although they are within 1 $\sigma$ of the other authors' models). For $V-I_{c}$, the derived ages for the three models are in close agreement, however, the \citeauthor{Baraffe1998} models again yield slightly older ages. Furthermore, \citeauthor{Baraffe1998} and \citeauthor{Siess2000} models yield derived DM that are larger than the \citeauthor{DAntona1997} $V-I_{c}$ models. 

Our results also show that all three models give systematically older ages for the $B-V$ CMDs compared to the $V-I_{c}$ ones. The color-dependent age discrepancies seen are of order $\sim$10 Myr, and are on the same order of offsets seen in other young open clusters when comparing isochrone fits using two different colors \citep[see][]{Naylor2002}. We also note that for the \citeauthor{Baraffe1998} and \citeauthor{Siess2000} models, there is a color-dependent offset of $\sim$0.2 magnitudes seen in the measured DM from the $V-I_{c}$ CMDs. We find the \citeauthor{DAntona1997} models provide the most consistent age and DM between the IC~4665 $B-V$ and $V-I_{c}$ CMDs. 

At this point, we caution the reader that the observed model- and color-dependent discrepancies in ages and DM are likely due to many different influences, and may not necessarily be representative of problems with low-mass stellar evolution models. Indeed, these three models include different physics that almost certainly must lead to different predicted luminosities and effective temperatures at given ages \citep[][for detailed discussion, see]{Siess2000,Hillenbrand2004}; however, the observed offsets may also be the result of systematics in the modeling process. For example, the ages derived from the $B-V$ CMDs may be older due to the $B-V$ color of late-type stars becoming nearly insensitive to changes in effective temperature, which consequentially makes it difficult to assign a PMS isochrone to such low-mass stars. In addition, the {\em Pleiades tuning} method may also introduce systematic errors into our analysis. For instance, \citet{Stauffer2007} find that low-mass stellar evolution models have trouble fitting the observed optical and infrared colors of low-mass members of the Pleiades. Because we calibrate our IC~4665 colors and magnitudes using an empirical fit to the Pleiades main-sequence, any such systematics in the Pleiades tuning method are included in our isochrone modeling procedure. More generally however, by employing the {\em Pleiades tuning} method we are assuming that the colors and magnitudes of the Pleiades main-sequence stars are representative of the younger stars in IC~4665. Unfortunately, our IC4665 dataset is relatively small compared to other well studied open clusters and does not include a complete, contaminant-free PMS, which therefore may not allow us to disentangle {\em all} of the different color- and model-dependent systematics affecting PMS isochrones modeling. 

Keeping this in mind, we would nevertheless still like to derive a single, best-fit PMS age and distance for IC~4665 for comparison purposes. Since the \citeauthor{Jeffries2009} catalog is most likely to contain the highest number of {\em bona fide} IC~4665 members, we use this catalog's resulting best-fit isochrone ages and DM as representative of the cluster. First, we calculate the inter-model error-weighted mean age and DM for the $B-V$ and $V-I_{c}$ colors, separately. The errors on these values are asymmetric, therefore the weights for our means are based on whether the age/DM lies above or below the straight average, {\em e.g.,} if an age lies below the straight average, we then weight that data point using its upper error value. Using this approach, we find color-dependent ages of 30.6$\pm$3.5 and 49.3$\pm$9.0 Myr, as well as intrinsic DM of 7.88$\pm$0.05 and 7.75$\pm$0.01 mag (377$\pm$9 and 355$\pm$3 pc), for our $B-V$ and $V-I_{c}$ CMDs, respectively. Taking the weighted mean of these color-dependent values, we get a final PMS age and DM for IC~4665 of 35.8$\pm$9.3 Myr and 7.78$\pm$0.07 mag (360$\pm$12 pc), for which the quoted errors are uncertainties in the mean. 

\begin{figure*} 
  \centering
  \includegraphics[viewport= 50 0 500 300,scale=1.0,angle=0]{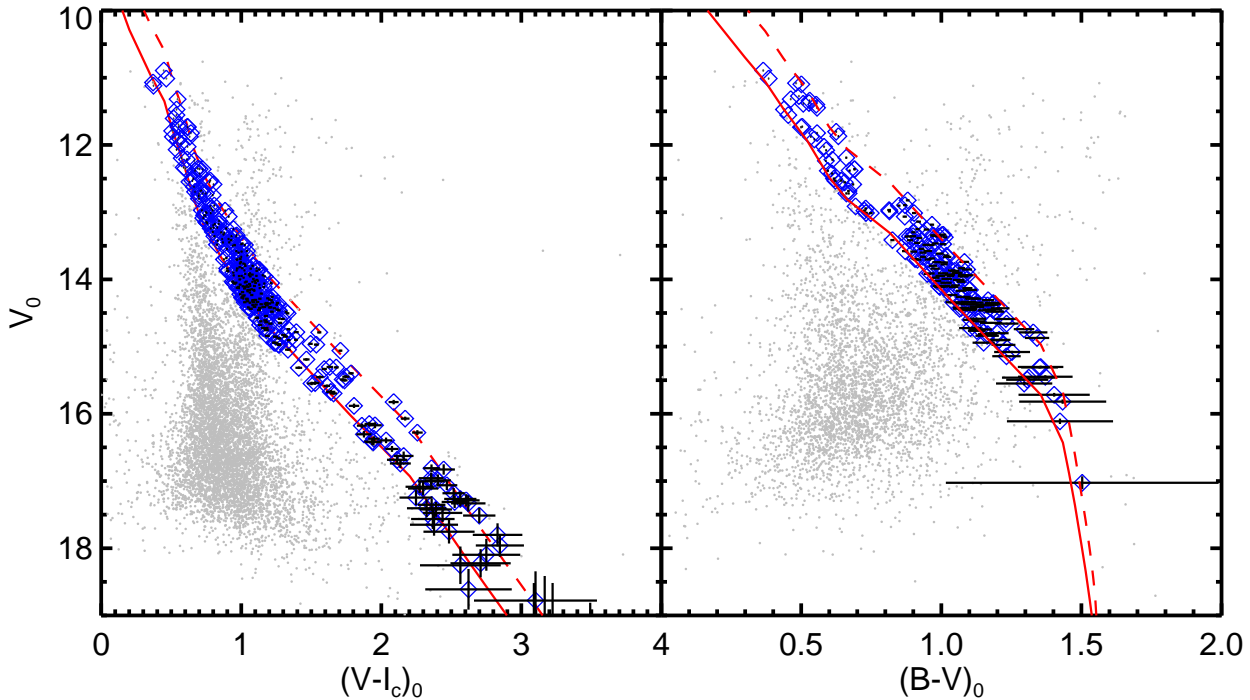} 
  \caption{
    \label{fig.finaliso}  
    Intrinsic $V-I_{c}$ and $B-V$ color-magnitude diagrams for IC~4665 are plotted 
    for our entire photometric catalog. Stellar objects having colors and 
    magnitudes consistent, within their photometric errors, of a 36 Myr and 
    360 pc \citeauthor{DAntona1997} model isochrone (red solid line) are 
    highlighted (blue diamonds), together with the locus of its associated 
    equal-mass binary sequence (red dotted line). Photometric data have been 
    dereddened using an $A_{V} = 0.543$, an $E(B-V) = 0.174$ and an $E(V-I_{c}) = 0.2175$. 
  }
\end{figure*}

The entire photometric catalog of IC~4665 fields is plotted in Fig.~\ref{fig.finaliso} in the form of $V/V-I_{c}$ and $V/B-V$ CMDs, where stars having photometric data consistent (within the errors) of being photometric members of 36 Myr and 360 pc \citeauthor{DAntona1997} isochrones are highlighted. For each of these candidate photometric members of IC~4665, we list their astrometric and photometric properties in Table \ref{tab.isomem}. This photometric membership selection is based upon \citeauthor{DAntona1997} because our model- and color-averaged age and DM are most closely matched to the values we get from the individual modeling of the $V-I_{c}$ and $B-V$ CMDs compared to the other theoretical PMS models generated by other authors ({\em e.g.,} \citeauthor{Baraffe1998} \& \citeauthor{Siess2000}). Although this membership list most likely contains a high number of cluster members, its exclusive use in identifying members of IC~4665 is not recommended due to the non-zero probability of field star contamination in the catalog, especially where the Galactic field star distribution intersects the cluster main-sequence close to $(V-I_{c})_{0} \sim 1.00$ and $(B-V)_{0} \sim 1.25$.

\input{tab6.tex}

Previously, we have shown that the CMDs of IC~4665 members appear very similar to those of the $\sim$35 Myr old NGC~2547 cluster (see Fig.~\ref{fig.isocomp}), which is now empirically supported by our PMS age determination for IC~4665. Moreover, the best-fit DM that we derive for the cluster is in excellent agreement with the trigonometric parallax distance resulting from the recent re-reduction of the HIPPARCOS astrometric dataset by \citet{vanLeeuwen2009}, which yields an intrinsic DM of 7.75$\pm$0.21 mag (355$\pm$35 parsecs).

\subsection{Upper-Main-Sequence Turn-off Age}\label{sec.highmass}

Comparing the positions of open cluster upper-main-sequences in color-magnitude space, \citet{Mermilliod1981a} determines that IC~4665 had a nuclear turn-off age of 36 Myr. However, P93 suggest that the upper main-sequence for IC~4665 is best comparable to that of the $\sim$70 Myr open cluster $\alpha$ Persei, and if effects due to rapid rotation are taken into account, the upper main-sequence is most like the Pleiades CMD, suggestive of a cluster age near to $\sim$100 Myr.

\begin{figure} 
  \centering
  \includegraphics[scale=0.9,angle=0]{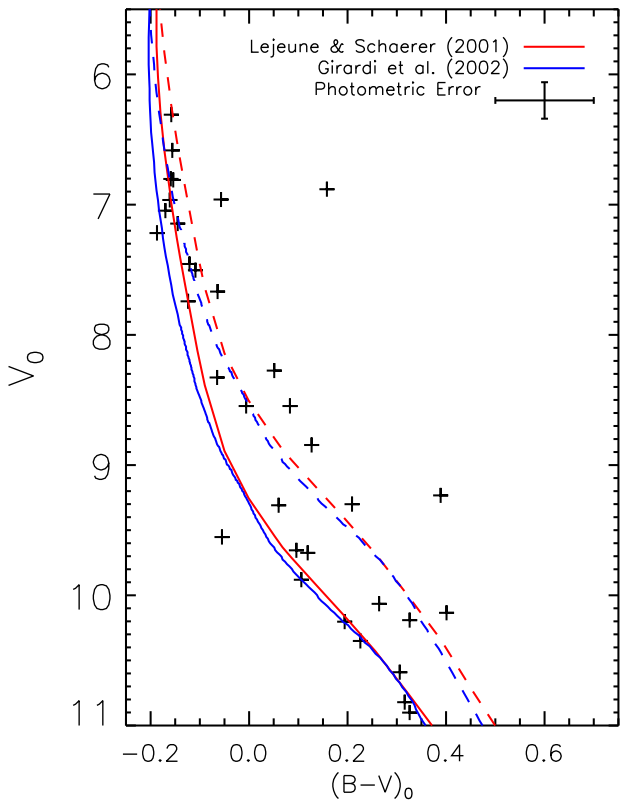} 
  \caption{
    \label{fig.highmass}
    High-mass (V$<$11 mag) color-magnitude diagram using photometric data from 
    the optical catalogs of \citet{Prosser1993} and \citet{Menzies1996}. Also 
    plotted are the best-fit isochrones (solid lines) from our upper-main-sequence 
    $\tau^{2}$-modeling using the two high-mass stellar models detailed in the 
    legend, as well as their associated equal-mass binary sequences (dotted lines). 
    The adopted uncertainty for the optical data is shown.
    }
\end{figure}

In light of this disagreement in nuclear turn-off age for IC~4665, we now use the $\tau^{2}$ technique, outlined in Section \ref{sec.lowmass}, to measure an age and distance for only the high-mass cluster members. Because the main-sequence turn-off for IC~4665 is located at a V magnitude brighter than our CTIO photometric survey saturation limit, we utilize the bright ($V<11$ magnitude) optical data from P93 and \citet{Menzies1996}. Unfortunately, neither catalog contained published individual uncertainties on their photometric data; therefore, we set the error in the $B$ and $V$ magnitudes to the level of observed scatter in P93 data in stars with $V<14$ mag found in our catalog comparison ($\Delta V=0.07$ mag and $\Delta B-V=0.1$ mag, see Section \ref{sec.comp}). Following \citet{Naylor2009}, we use upper-main-sequence models from \citet{Lejeune2001} and \citet{Girardi2002}, coupled with solar-metallicity bolometric corrections and effective temperature to color conversions, as well as color dependent reddening vectors from \citet{Bessell1998}. We find upper-main-sequence ages for IC~4665 of 41$\pm$12 and 42$\pm$12 Myr and intrinsic DM of 7.76$\pm$0.04 and 7.77$\pm$0.04 mag (356$\pm$10 and 358$\pm$15 parsecs) for the aforementioned different high-mass stellar evolution models, respectively. We plot in Fig.~\ref{fig.highmass} the photometric data used in our high-mass CMD modeling, as well as our two best-fit isochrones. We note that this age measurement is not thoroughly constrained due to the lack of high-mass cluster members tracing out IC~4665's upper-main-sequence; nevertheless, we find that the nuclear turn-off age for IC~4665 we have calculated is similar to the previous estimate from \citet{Mermilliod1981a}, and within the error budget, are in agreement with the $\alpha$ Persei-like age estimate from P93 (although quite discrepant from P93's Pleiades-like age estimate). 

\citet{Naylor2009} recently investigated the differences observed in isochronal ages measured from upper- and lower-main-sequences of young ($<$ 100 Myr) open clusters. They find a trend of PMS {\em turn-on} ages being $\sim$1.5 times smaller than nuclear {\em turn-off} ages for a selection of young open clusters. Interestingly, our current study on IC~4665 shows the upper-main-sequence isochrone models have best-fit ages that are similar (difference of $\sim$6 Myr) to the PMS isochrone age while using the $B-V$ low-mass CMD. However, if we use the PMS isochrone age from the $V-I_{c}$ CMDs (30.6$\pm$3.5 Myr) or our inter-model $V-I_{c}$ mean PMS age (35.8$\pm$9 Myr), we find $\sim$1.5 times these PMS ages are formally within uncertainty of the nuclear {\em turn-off} age ({\em i.e.}, $1.5\times30.6\pm3.5$ Myr $= 45.9\pm3.5$ Myr and $1.5\times35.8\pm9$ Myr $= 53.7\pm9$ Myr, compared to our {\em turn-off} age of $42\pm12$ Myr), corroborating the \citet{Naylor2009} findings. 

\subsection{Lithium Depletion Boundary Age}\label{sec.LDBage}

Using the lithium depletion boundary to date young open clusters provides a precise ($\sim10\%$ uncertainty), nearly model-insensitive method to determine ages for open clusters \citep{Burke2004,Jeffries2005}. It has been found that LDB ages for open clusters tend to be systematically older than upper-main-sequence {\em turn-off} ages \citep{Stauffer1998,Stauffer1999,BarradoyNavascues2004,Burke2004}; however, the discrepancy between LDB and {\em turn-off} ages has been found to be smaller for younger open clusters \citep{Jeffries2005,Naylor2009} and may not exist for PMS isochrone ages \citep{Jeffries2005}.

IC~4665 provides an interesting test for these age discrepancies in young open clusters because it is one of the few clusters with a LDB age in addition to well-defined PMS and upper-main-sequence populations. We recall that, \citet{Manzi2008} measured a LDB age of 28$\pm$5 Myr age for IC~4665, which is in agreement within the error budget, with our 35.8$\pm$9.3 Myr model-averaged PMS age determination. We also find the upper-main-sequence age for IC~4665 (42$\pm$12 Myr) is 1.5 times {\em older} than the LDB age, thereby contradicting the trend seen in other open clusters with LDB ages. We note, however, that the paucity of high-mass stars in PMS open clusters like IC~4665 reduces the fidelity of their {\em turn-off} ages, therefore this discrepancy in the observed {\em turn-off}-LDB age trend may turn out to be of little significance. 

\subsection{X-ray Activity and Age}\label{sec.xray} 

\begin{figure*} 
  \centering
  \includegraphics[viewport= 170 0 700 500,scale=0.75,angle=90]{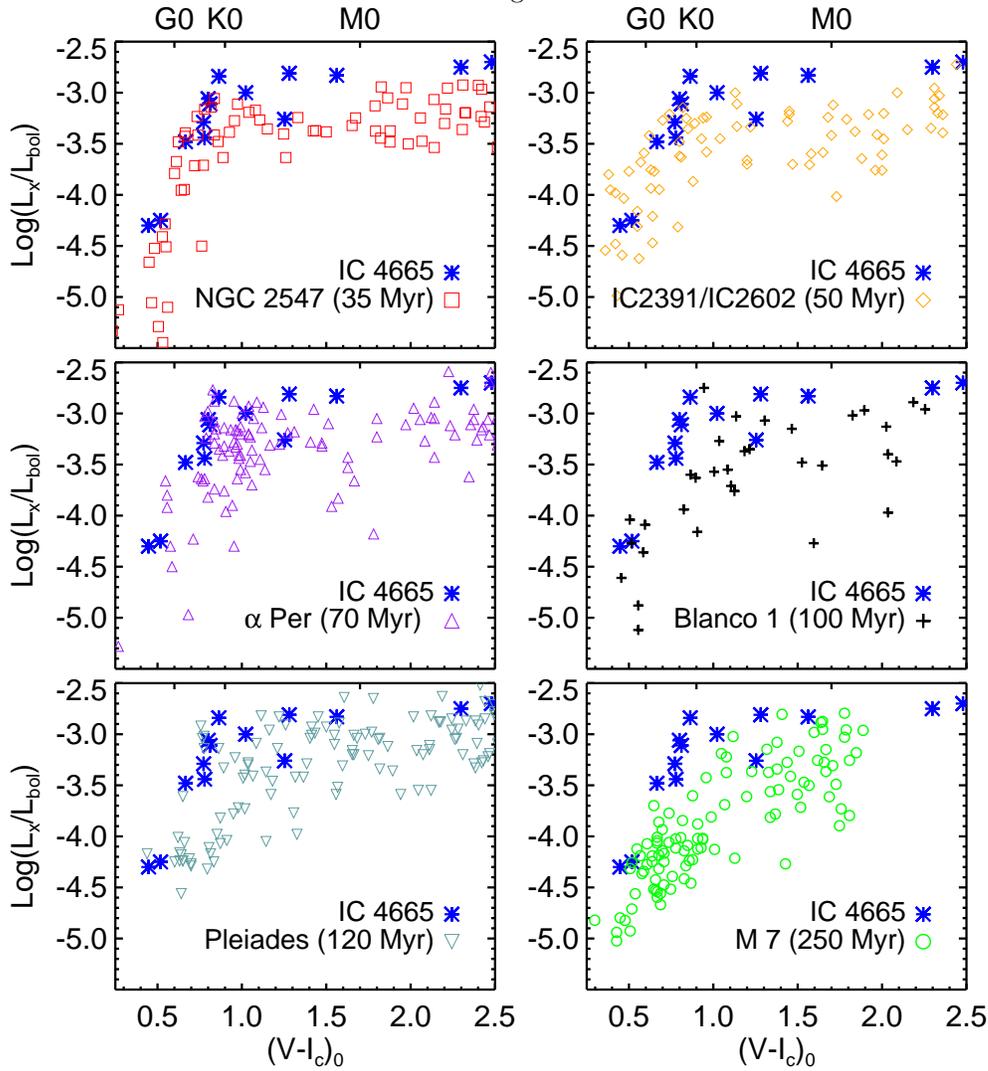} 
  \caption{
    \label{fig.xray}  
    Ratio of X-ray to bolometric luminosity plotted as a function of intrinsic 
    $V-I_{c}$ color for X-ray sources taken from \citet{Giampapa1998} with 
    counterparts in our optical catalog for IC~4665 (blue asterisks). In each relevant 
    panel, also plotted are X-ray data for NGC~2547 (red squares), IC~2391/IC~2602 
    (orange diamonds), $\alpha$ Persei (purple triangles), Blanco~1 (black crosses), 
    Pleiades (light blue triangles), and M~7 (green circles). Data 
    are taken from \citet{Jeffries2006}, \citet{Stauffer1997}, \citet{Randich1996}, 
    \citet{Cargile2009}, \citet{Stauffer1994}, and \citet{Prosser1995b}, respectively. 
    Spectral-type ranges, as defined by \citet{Bessell1998}, are plotted along the top axes. 
    Intrinsic colors were calculated using $E(B-V)$ values collated in the WEBDA database 
    and an $E(V-I_{c}) = 1.25 \times E(B-V)$ relation.
    }
\end{figure*}

There is a very distinct, age- and mass-dependent morphology of X-ray activity observed in young open clusters that is fundamentally tied to the causal relationship between age-dependent stellar rotation and the presence of a rotationally-induced, magneto-hydrodynamic dynamo in young stars \citep[for review, see][]{Guedel2004}. More specifically, the distribution of X-ray emission, typically expressed as the distance-independent ratio of X-ray to bolometric luminosity ($L_{x}/L_{bol}$), can be described by two regimes: the first, a regime where $L_{x}/L_{bol}$ more or less monotonically increases with decreasing stellar mass (and independently, with decreasing stellar period); and second, the so called saturated regime, where there is a plateau in X-ray emission at $L_{x}/L_{bol} \sim 10^{-3}$ regardless of changes in stellar mass or change in stellar rotation rate \citep{Hempelmann1995,James2000}. The physical cause of this saturated regime is still a matter of debate \citep{Charbonneau1992,Odell1995,Jardine1999}; however, the transition point between these two regimes is empirically found to be a function of rotation period for a given stellar mass \citep{Pizzolato2003}, and therefore, a function of age due to the time-dependence of stellar rotation \citep{Barnes2003a,Barnes2007,Mamajek2008}. For example, in a young cluster like NGC~2547 ($\sim$35 Myr), even the higher mass stars ($\sim$ G-type stars) are rotating significantly fast enough to be in the saturated X-ray regime \citep{Jeffries2000}, while in an older cluster like the Hyades ($\sim$700 Myr) the transition point between the two regimes is found in the M dwarfs mass domain \citep{Stern1995}. Consequently, we can assign approximate ages of open clusters, in a relative manner, by comparing their X-ray distributions to those of clusters with known ages, and comparing the observed color ({\em i.e.,} stellar mass) at which there is a transition from the linearly increasing, color-dependent X-ray luminosity function to the saturated regime. 

In the ROSAT study of IC~4665 \citep{Giampapa1998}, the authors suggest that IC~4665 has a X-ray distribution indicative of a cluster which is older than the $\alpha$ Persei ($\sim$70 Myr), and very similar to the Pleiades ($\sim$120 Myr). They find that the color at which the IC~4665 X-ray distribution appears to transition to the saturated regime is redder (lower-mass) than $\alpha$ Persei, and is in agreement with the color of saturation for the Pleiades. Therefore, according to the age-activity relationship, IC~4665 should have an age of $\sim$120 Myr. This finding is formerly discordant with our current high- and low-mass isochronal analysis of IC~4665 (Sections \ref{sec.lowmass} and \ref{sec.highmass}), as well as its measured LDB age.

\begin{figure*} 
  \centering
  \mbox{\subfigure{\includegraphics[width=3.25in]{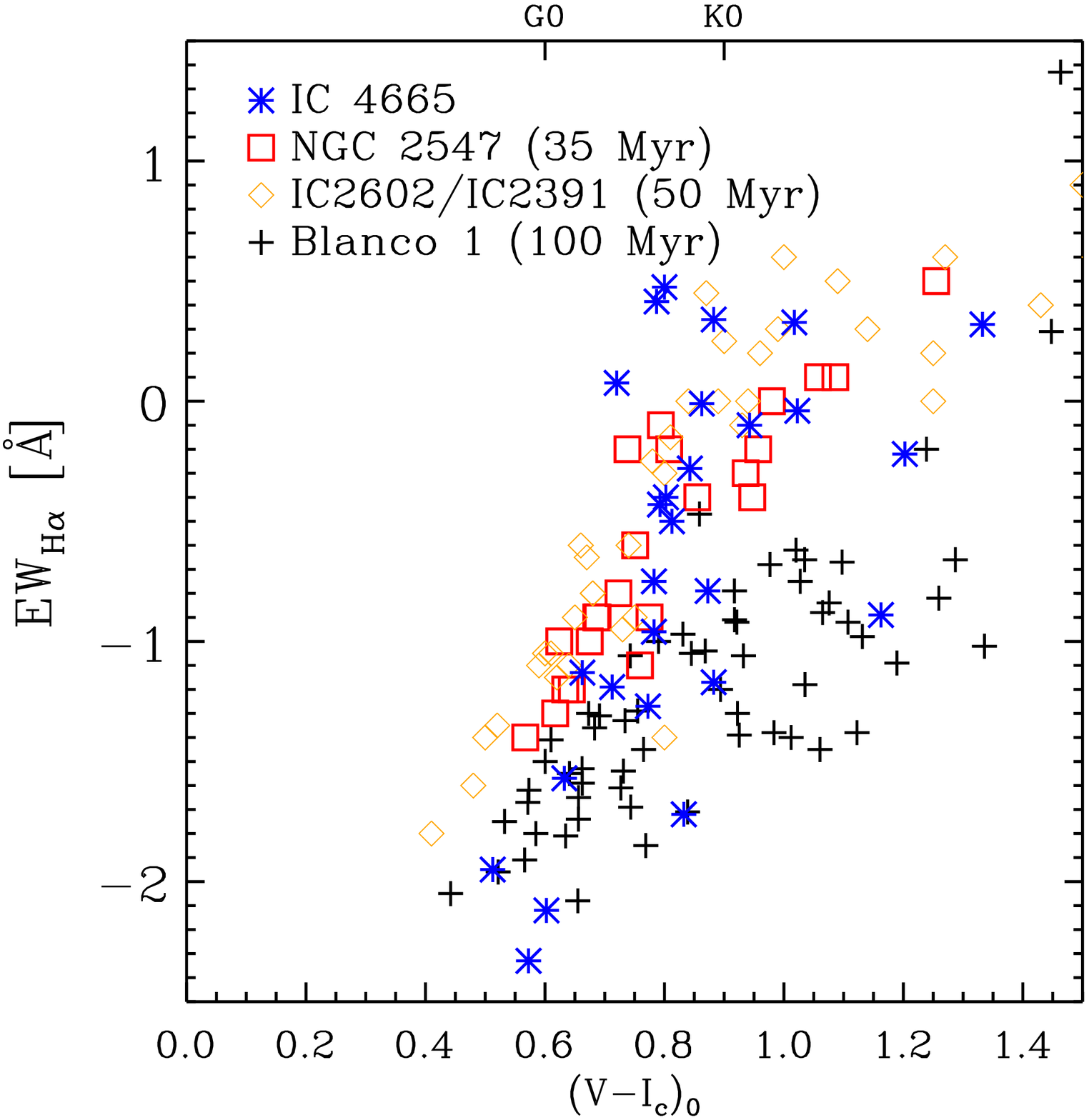} 
        \subfigure{\includegraphics[width=3.25in]{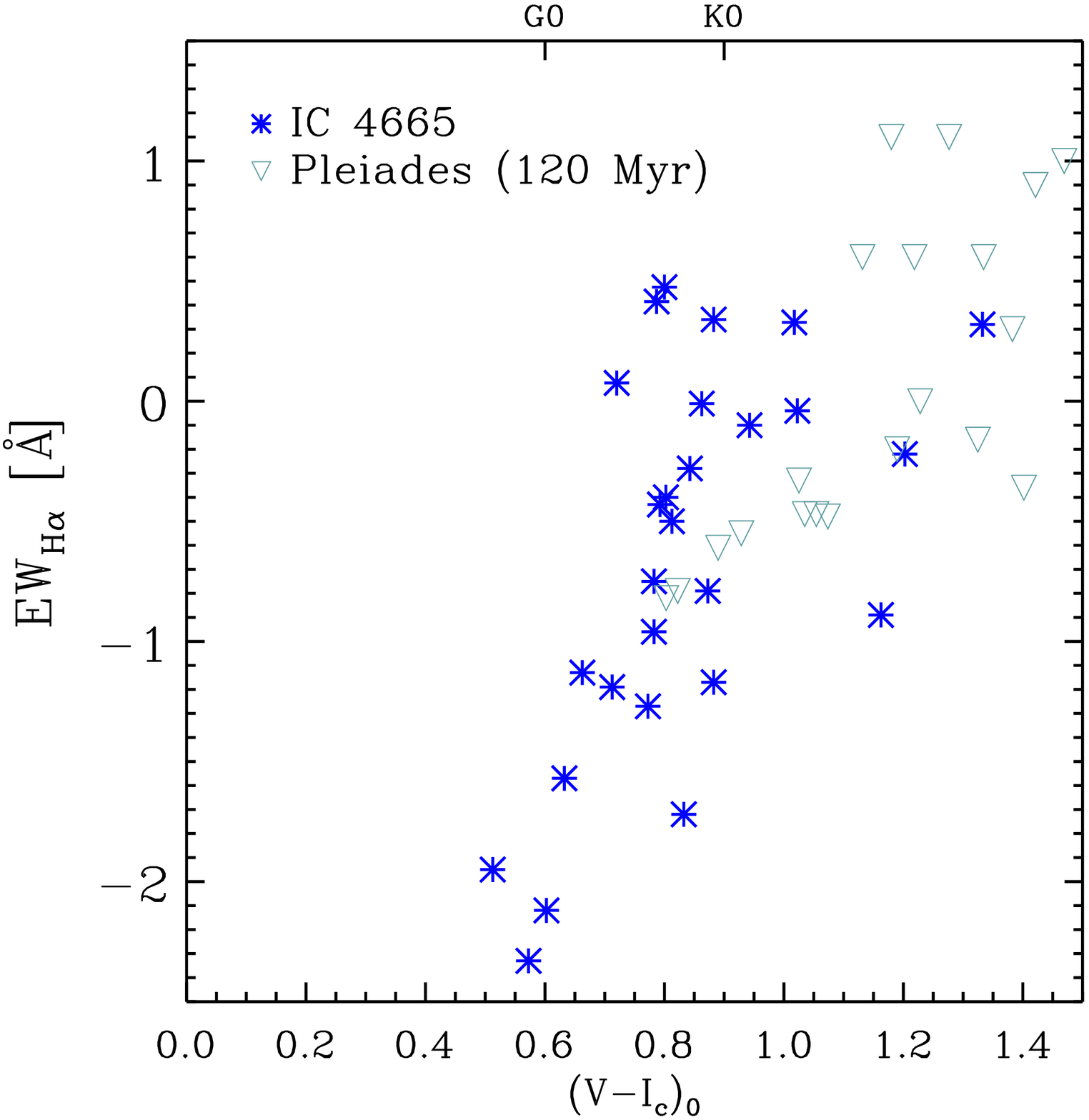} }}}
  \caption{
    Equivalent width (EW) of the Balmer series H$\alpha$ spectral line is plotted versus 
    intrinsic $V-I_{c}$ color for matched optical counterparts in our 
    IC~4665 catalog (blue asterisks), as well as data for (left panel) NGC~2547 
    (red squares), IC~2391/IC~2602 (orange diamonds), Blanco~1 
    (black crosses), and (right panel) the Pleiades. H$\alpha$ data are taken from 
    \citet{Prosser1993,Martin1997,Jeffries2009}, \citet{Jeffries2000}, \citet{Stauffer1997},
    \citet{Panagi1997}, and \citet{Stauffer1987} for each cluster, respectively. 
    Spectral-type ranges, as defined by \citet{Bessell1998}, are plotted along the top axes.
    We deredden the $V-I_{c}$ colors using the cluster reddening coefficients listed in WEBDA, and 
    defined positive $EW_{H\alpha}$ to be in emission.
  } 
  \label{fig.halpha}  
\end{figure*} 

In order to evaluate the validity of this X-ray age determination, we utilize our new optical photometry and several X-ray surveys of other clusters published since the \citet{Giampapa1998} study, including several with data from the new generation of X-ray telescopes (Chandra and XMM-Newton X-ray observatories). For ROSAT X-ray detected stars in IC~4665, we assign new colors ($V-I_{c}$) and calculate new bolometric luminosities from our optical catalog. We use the bolometric corrections from \citet{Johnson1966} for stars with intrinsic $V-I_{c} < 1.6$ and the formalism given in \citet{Monet1992} for stars with intrinsic $V-I_{c} > 1.6$. In Fig.~\ref{fig.xray}, we show our new $L_{x}/L_{bol}$ values as a function of intrinsic $V-I_{c}$ color for IC~4665, as well as the X-ray distribution of the 35 Myr NGC~2547 \citep{Jeffries2006}, 50 Myrs IC~2391/IC~2602 \citep{Stauffer1997}, 70 Myr $\alpha$~Persei, 100 Myr Blanco~1 \citep{Cargile2009}, 120 Myr Pleiades \citep{Stauffer1994}, and 250 Myr M~7 \citep{Prosser1995b} open clusters. 

Looking at the overall morphology of X-ray emission in Fig.\ \ref{fig.xray}, we find that the X-ray distribution of IC~4665, and more specifically the color at which saturated stars are observed, is much more in agreement with the young NGC~2547 cluster than the older open clusters. According to the relationship between age and activity, the similarity of the two cluster's X-ray activity distributions suggests that IC~4665 has an age nearer to $\sim$35 Myr \citep[][]{Jeffries2005}, thus is in agreement with our isochrone ages and the published LDB age.

Although the IC~4665 data include our new SMARTS optical photometry, the onset of saturation in $\alpha$ Persei and the Pleiades is still observed to be very similar to IC~4665 (and furthermore NGC~2547). In light of our larger cluster sample compared to \citeauthor{Giampapa1998}, we find that the X-ray emission distribution of the Pleiades and $\alpha$ Persei appears to be in disagreement with the overall observed activity-age morphology. We note that \citet{Cargile2009} similarly find the Pleiades X-ray distribution is at odds with the approximately coeval open cluster Blanco~1, and speculate that this discrepancy is primarily due to scatter in the Pleiades photometric data, and is not attributable to the stellar properties of Blanco~1 members.

\subsection{H$\alpha$ emission and Age}\label{sec.halpha}

Much like X-ray activity, studies of chromospheric activity have also been used as a good indicator of relative age for open clusters \citep{Prosser1991,Stauffer1991,Prosser1992}. Through its causal relationship to the solar-type dynamo, the level of chromospheric emission changes as a function of stellar rotation, and therefore ultimately on age \citep{Mamajek2008}. More specifically, as a star spins down with age, reduced emission is observed in chromospheric lines, {\em e.g.,} H$\alpha$, \ion{Ca}{2} H \& K \citep{Wilson1963}.

Similar to the X-ray study of \citeauthor{Giampapa1998}, P93 claim that IC~4665 has an age of $\gtrsim$100 Myr based on the upper envelope of its H$\alpha$ equivalent width distribution, as a function of pseudo-MK (pMK) spectral-type, being below that seen for the Pleiades, thus in direct conflict with our current isochrone dating analysis and its measured LDB age. P93 derived pMK spectral-types using a series of line ratios, mostly strong TiO features, measured from their low-dispersion spectra. Here, using our new photometric catalog and H$\alpha$ data listed in Section \ref{sec.lowmass}, we reanalyze the H$\alpha$ distribution in IC~4665 to investigate the reported chromospheric activity-age estimate of $\sim$100 Myr. In Fig.~\ref{fig.halpha} left panel, we plot H$\alpha$ equivalent width (EW$_{H\alpha}$) taken from P93 as a function of our new intrinsic $V-I_{c}$ color. We also plot H$\alpha$ data for open clusters NGC~2547 \citep[35 Myr;][]{Jeffries2000}, IC~2391/IC~2602 \citep[50 Myr;][]{Stauffer1997} and Blanco 1 \citep[100 Myr;][James et al. in prep]{Panagi1997}. The level of H$\alpha$ emission seen, as well as the color at which H$\alpha$ is observed in emission, in IC~4665 indicates that the cluster is younger than 50 Myr, and most likely has an age near (or younger) than that of NGC~2547 (35 Myr). 

When we plot the H$\alpha$ emission distribution of IC~4665 and the Pleiades \citep{Stauffer1987} using $V-I_{c}$, instead of using the pMK spectral-type classification schema, as a proxy for effective temperature (Fig.~\ref{fig.halpha} right panel), we see that the two clusters have dissimilar morphologies. The onset of H$\alpha$ in emission in IC~4665 is found at an earlier spectral-type compared to the Pleiades, thus suggesting that IC~4665 is younger than the $\sim$120 Myr old Pleiades. This finding is contrary to the chromospheric dating analysis in P93; however, their H$\alpha$ line measurements were performed over a much more limited range of spectral-types with the earliest being $\sim$M0. With our more extensive sample of H$\alpha$ measurements from three different sources \citep{Prosser1993,Martin1997,Jeffries2009}, we clearly observe H$\alpha$ emission in IC~4665 stars as early as spectral-type $\sim$G0 ($(V-I_{c})_{0} \sim 0.7$), thereby suggesting a much earlier age estimation.

\section{Summary \& Concluding Remarks}\label{sec.summary}
In this paper, we present the results of a new, homogeneous, standardized BVI$_{c}$ photometric survey of the central region of the young open cluster IC~4665. This dataset improves upon the previous work of P93, by incorporating a high-fidelity standardization procedure involving observations of multiple standards taken contemporaneously with our IC~4665 data. Our full photometric catalog contains 7359 detected point sources, of which all but 452 are matched within 3.0 arcseconds of 2MASS sources. The internal stability of our photometric system, as determined by a nightly control field, is $3-4\%$ for objects with B, V, and I$_{c}$ $<$ 16 magnitudes. Using our best-fit age and distance from our isochrone modeling (see \ref{sec.pmsresults}) we have identified 382 IC~4665 photometric cluster member from our photometry; 324 and 176 from the $V-I_{c}$ and $B-V$ CMDs, respectively. We provide this extensive membership catalog due to it likely containing a large number of cluster members, we advise that it not be used exclusive to identify clusters members due to a high level of field star contamination in IC~4665. 

In order to refine our photometric membership catalog into a high-fidelity membership list for IC~4665, we correlate the source positions in our new BVI$_{c}$ photometric survey with those in various extant membership catalogs. This includes membership based on kinematic properties, lithium absorption, rapid rotation, and chromospheric/coronal emission. We employ these membership catalogs with our updated photometry to reevaluate the age of IC~4665 using various dating techniques. New pre-main-sequence {\em turn-on} and upper-main-sequence {\em turn-off} ages were calculated using $\tau^{2}$ modeling, a least-squares fitting procedure well suited for modeling color-magnitude diagrams due to it taking into account correlated errors in age and distance. We find a best-fit PMS isochrone age and distance of 35.8$\pm$9 Myr and 360$\pm$12 pc, and a {\em turn-off} age and distance of 42$\pm$12 Myr and 357$\pm$13 pc, where these values are weighted averages and the errors are uncertainties on the means. 

Definite model- and color-dependent offsets are seen over a range of $\sim$10-20 Myr in age and $\sim$20 parsecs in distance, particularly in the results from the PMS isochrones. This large spread in ages and distances, as well as the systematic offsets seen between colors and models, are likely due to limitations in the individual membership catalogs ({\em e.g.,} a narrow range of colors for a given membership catalog produces large uncertainties in its derived age and distance) and/or systematic differences in the predicted stellar parameters from the different sets of models. At this point, how these systematic model- and color-dependent offsets influence the measured ages and distances cannot be completely deciphered primarily due to a limited number of known clusters members and a relatively high non-member contamination level of IC~4665. 

In addition, we also compared the distribution of magnetic activity in IC~4665 stars to other well studied open clusters. The coronal X-ray and chromospheric H$\alpha$ emission observed in IC~4665 is most similar to 30-40 Myr old open clusters ({\em e.g.,} NGC~2547), thereby providing further confirmation on our measured {\em turn-on} and {\em turn-off} ages. This result also provides further evidence against previous claims that IC~4665 has an observed activity distribution similar to the Pleiades. Previous IC~4665 activity-age studies have been reliant on relative distribution comparisons with clusters that we find do not fit into the observed global trends seen in larger samples of open clusters, particularly the X-ray emission distribution of $\alpha$ Persei and the Pleiades.

While it is clear that we do see slight offsets and differences in the various dating techniques we have employed in this paper, the majority of our measured ages for IC~4665 are consistently between $\sim$30-45 Myr. Interestingly, the lack of large systematic differences (factors of $>$1.5 in age) between dating techniques is not typical for open clusters; for example, published ages for the Pleiades range from 70 Myr \citep{Mermilliod1981b} to 150 Myr \citep{Burke2004}. Understanding why IC~4665 does not suffer from these systematic offsets should allow for a critical comparison between the different methods available to determine stellar ages. 

\acknowledgments
We recognize support for P.A.C. and D.J.J. from the National Science Foundation Career Grant AST-0349075 (P.I. Stassun, K.~G.). We cordially thank T.~Naylor for his helpful suggestions and guidance with the $\tau^{2}$ software. We also gratefully acknowledge the staff at CTIO and the SMARTS Consortium. The Cerro Tololo Inter-American Observatory, and the National Optical Astronomy Observatory, are operated by the Association of Universities for Research in Astronomy, under contract with the National Science Foundation. This research has made use of the SIMBAD database, operated at CDS, Strasbourg, France. This publication makes use of data products from the Two Micron All Sky Survey, which is a joint project of the University of Massachusetts and the Infrared Processing and Analysis Center/California Institute of Technology, funded by the National Aeronautics and Space Administration and the National Science Foundation.

\end{document}

%% file: tab1.tex
\begin{deluxetable}{l c c r}
\tablecolumns{4}
\tablewidth{0pc}
\tablecaption{Summary of SMARTS 1.0m Observations of IC~4665
\label{tab.pointings}}
\tablehead{
\colhead{IC~4665}                     &
\colhead{R.A.\tablenotemark{a}}       & 
\colhead{Dec\tablenotemark{a}}        & 
\colhead{Date of}                     \\
\colhead{Field\tablenotemark{b}}      &
\colhead{[HH:MM:SS]}                  & 
\colhead{[DD:MM:SS]}                  & 
\colhead{Observation}}
\startdata
F1 &   17:47:29.8 & +06:01:00.0 & 16/Sept/2005 \\
F2 &   17:46:13.5 & +06:01:00.0 & 16/Sept/2005 \\
F3 &   17:44:56.3 & +06:01:00.0 & 17/Sept/2005 \\
F4 &   17:47:29.8 & +05:41:52.0 & 15/Sept/2005 \\
F5 &   17:46:13.5 & +05:41:52.0 & 15/Sept/2005 \\
F6 &   17:44:56.3 & +05:41:52.0 & 17/Sept/2005 \\
F7 &   17:47:29.8 & +05:22:56.0 & 15/Sept/2005 \\
F8 &   17:46:13.5 & +05:22:56.0 & 16/Sept/2005 \\
F9 &   17:44:56.3 & +05:22:56.0 & 16/Sept/2005 \\
CF &   17:46:52.6 & +05:32:00.0 & 15/Sept/2005 \\
CF &   17:46:52.6 & +05:32:00.0 & 17/Sept/2005 \\
\enddata
\tablenotetext{a}{Coordinates are Y4KCam field centers. All coordinates are J2000.0 epoch.}
\tablenotetext{b}{F'n'-fields represent the main survey 
area of the cluster in a $3\times3$ mosaic covering about 
one square degree centered close to the cluster's WEBDA 
coordinates.  Photometric data for a control field (CF) 
were also procured to allow us to assess the photometric 
integrity of the dataset.}
\end{deluxetable}

%% file: tab2.tex
\begin{deluxetable*}{clccrccc}
\tablecolumns{8}
\tablewidth{0pc}
\tablecaption{Nightly Extinction, Transformation 
and Zero-point Coefficients
\label{tab.trans}}
\tablehead{\colhead{Date}             &  
\colhead{Equation}                    &
\colhead{Extinction\tablenotemark{a}} & 
\colhead{CTC\tablenotemark{b}}        & 
\colhead{Zero}                        & 
\colhead{Mean\tablenotemark{c}}       & 
\colhead{RMS\tablenotemark{d}}        & 
\colhead{No. of\tablenotemark{e}}    \\
\colhead{}                            &
\colhead{}                            &
\colhead{Coefficient}                 & 
\colhead{}                            & 
\colhead{Point}                       &
\colhead{$\Delta$ mag}                 &
\colhead{$\Delta$ mag}                 &
\colhead{Standards}}
\startdata
15/Sept/2005   & V (B-V)  & 0.1478 & 0.0140 & -1.8654 & $ 6.3 \times 10^{-05}$ & 0.0148 & 90, 72 \\
               & V (V-I$_{c}$) & 0.1560 & 0.0128 & -1.8541 & $ 1.0 \times 10^{-05}$ & 0.0176 & 87, 75 \\
               & B-V      & 0.1116 & 0.8940 & -0.3011 & $ 4.5 \times 10^{-05}$ & 0.0166 & 90, 71 \\
               & V-I$_{c}$ & 0.1235 & 1.0009 &  1.0472 & $ 1.0 \times 10^{-05}$ & 0.0172 & 89, 58 \\
16/Sept/2005   & V (B-V)  & 0.1476 & 0.0166 & -1.8518 & $ 2.4 \times 10^{-05}$ & 0.0157 & 80, 62 \\
               & V (V-I$_{c}$) & 0.1492 & 0.0171 & -1.8524 & $ 1.0 \times 10^{-05}$ & 0.0170 & 74, 62 \\
               & B-V      & 0.1497 & 0.8731 & -0.2334 & $ 7.5 \times 10^{-05}$ & 0.0177 & 80, 60 \\
               & V-I$_{c}$     & 0.0832 & 1.0001 &  0.9935 & $ 6.5 \times 10^{-05}$ & 0.0172 & 80, 60 \\
17/Sept/2005   & V (B-V)  & 0.1556 & 0.0272 & -1.8307 & $ 8.1 \times 10^{-05}$ & 0.0180 & 76, 57 \\
               & V (V-I$_{c}$) & 0.1574 & 0.0239 & -1.8290 & $-5.0 \times 10^{-05}$ & 0.0183 & 76, 57 \\
               & B-V      & 0.0460 & 0.8808 & -0.3565 & $-7.7 \times 10^{-05}$ & 0.0220 & 64, 52 \\
               & V-I$_{c}$     & 0.0163 & 1.0040 &  0.9107 & $ 3.3 \times 10^{-05}$ & 0.0190 & 74, 54 \\
\enddata
\tablenotetext{a}{In units of magnitude/airmass.}
\tablenotetext{b}{Color Transformation Coefficient}
\tablenotetext{c}{Observed mean difference between calculated and published magnitudes or colors for 
\citet{Landolt1992,Landolt2009} standard stars.}
\tablenotetext{d}{Observed RMS difference between calculated and published magnitudes or colors for 
\citet{Landolt1992,Landolt2009} standard stars.}
\tablenotetext{e}{$m,n$ -- where $m$ is the number of Landolt standard stars observed, and $n$ 
represents the number of standard stars used in the final photometric solution.}
\end{deluxetable*}

%% file: tab3a.tex
\begin{deluxetable*}{l l c c c c c }
\tablecolumns{7}
\tablewidth{0pt}
\tabletypesize{\footnotesize}
\tablecaption{SMARTS BVI$_{c}$ and JHK$_{s}$ Photometry for the Region Around IC~4665\label{tab.full}}
\tablehead{
\colhead{Mosaic} & \colhead{Object} & 
\colhead{R.A.} & \colhead{Dec.} & 
\colhead{V} & \colhead{B-V} & \colhead{V-I$_{c}$} \\ 
\colhead{Field} & \colhead{Number} & 
\colhead{[HH:MM:SS]} & \colhead{[DD:MM:SS]} & 
\colhead{} & \colhead{} & \colhead{} 
}
\startdata
F6 & 682 & 17:44:17.47 & +05:38:34.2 & 18.299$\pm$0.193 & 99.999$\pm$9.999 & 1.566$\pm$0.243  \\
F6 & 259 & 17:44:17.59 & +05:40:44.7 & 15.961$\pm$0.022 &  1.083$\pm$0.063 & 1.232$\pm$0.033 \\
F6 & 723 & 17:44:17.62 & +05:36:05.0 & 17.051$\pm$0.061 & 99.999$\pm$9.999 & 1.006$\pm$0.101 \\
\enddata

\end{deluxetable*}

%% file: tab3b.tex
\begin{deluxetable*}{l l c c c c c c c}
\tablecolumns{9}
\tablewidth{0pt}
\setlength{\tabcolsep}{0.03in}
\tabletypesize{\footnotesize}
\tablenum{3}
\tablecaption{cont. }
\tablehead{
\colhead{Mosaic} & \colhead{Object} & 
\colhead{R.A.(2MASS)} & \colhead{Dec.(2MASS)} & 
\colhead{Pos. Off. \tablenotemark{a}} & 
\colhead{J} & \colhead{H} & \colhead{K$_{s}$} & \colhead{2MASS} \\
\colhead{Field} & \colhead{Number} & 
\colhead{[HH:MM:SS]} & \colhead{[DD:MM:SS]} & 
\colhead{[$\arcsec$]} & 
\colhead{} & \colhead{} & \colhead{} & \colhead{Qual Flag} 
}
\startdata
F6 & 682 & 17:44:17.41 & +05:38:34.60 & 0.96 & 16.149$\pm$0.108 & 15.819$\pm$0.167 & 15.470$\pm$0.209 & ACC \\
F6 & 259 & 17:44:17.55 & +05:40:44.52 & 0.60 & 13.965$\pm$0.030 & 13.457$\pm$0.036 & 13.325$\pm$0.040 & AAA \\
F6 & 723 & 17:44:17.59 & +05:36:04.49 & 0.62 & 15.437$\pm$0.071 & 15.027$\pm$0.088 & 15.181$\pm$0.161 & AAC \\
\enddata
\tablecomments{Only a few rows are displayed here to show the content of the table, the full table is available in electronic form. 
All positions are given in J2000 coordinates. Magnitude, colors, positions, and uncertainties given as 99.999 or 9.999 indicate that the
value is undetermined.}
\tablenotetext{a}{The positional offset between the R.A. and Dec. given in our photometric catalog and the corresponding 2MASS source position.}
\end{deluxetable*}

%% file: tab4.tex
\begin{deluxetable*}{c l c l c l c}
\tablecolumns{7}
\tablewidth{0pt}
\tablecaption{Internal Precision of SMARTS BVI$_{c}$ Photometry\label{tab.internal}}
\tablehead{
\colhead{ } & \multicolumn{2}{c}{15th - 17th} & \multicolumn{2}{c}{15th - 16th} &  \multicolumn{2}{c}{16th - 17th} \\
\colhead{Filter} & \colhead{Mean} & \colhead{Num} & \colhead{Mean} & \colhead{Num} & \colhead{Mean} & \colhead{Num} 
}
\startdata
Full Catalog \\
\hline
V       & 0.024$\pm$0.056 & 1106 & 0.019$\pm$0.062 & 234 & 0.030$\pm$0.066 & 206 \\
B       & 0.079$\pm$0.061 &  669 & 0.058$\pm$0.070 & 107 & 0.032$\pm$0.075 & 110 \\
I$_{c}$ & 0.029$\pm$0.071 & 1058 & 0.064$\pm$0.061 & 219 & 0.004$\pm$0.076 & 190 \\
\hline
V $<$ 16 \\
\hline
V       & 0.016$\pm$0.014 & 365 & 0.030$\pm$0.014 & 72 & 0.001$\pm$0.015 & 65  \\
B       & 0.066$\pm$0.038 & 166 & 0.054$\pm$0.040 & 31 & 0.020$\pm$0.045 & 30  \\
I$_{c}$ & 0.020$\pm$0.014 & 794 & 0.044$\pm$0.013 & 155 & 0.011$\pm$0.015 & 138 \\
\hline
V $>$ 16 \\
\hline
V       & 0.028$\pm$0.077 & 741 & 0.014$\pm$0.084 & 162 & 0.045$\pm$0.090 & 141 \\
B       & 0.092$\pm$0.124 & 503 & 0.055$\pm$0.137 &  76 & 0.070$\pm$0.146 &  80 \\
I$_{c}$ & 0.036$\pm$0.091 & 264 & 0.046$\pm$0.080 &  64 & 0.010$\pm$0.098 &  52  \\
\enddata
\end{deluxetable*}

%% file: tab5.tex
\begin{deluxetable*}{l l c c c c}
\tablecolumns{6}
\tablewidth{0pt}
\setlength{\tabcolsep}{0.2in}
\tablecaption{Isochrone Ages and Distances for IC~4665
\label{tab.isoresult}}
\tablehead{
\colhead{Model} &
\colhead{Parameter\tablenotemark{a}} & 
\multicolumn{2}{c}{Median\tablenotemark{b}} & 
\multicolumn{2}{c}{\citeauthor{Jeffries2009}} \\
\colhead{} &
\colhead{} & 
\colhead{$B-V$} & 
\colhead{$V-I_{c}$} & 
\colhead{$B-V$} & 
\colhead{$V-I_{c}$} 
}
\startdata
DAM97\tablenotemark{c} & Age~[Myr]  &  48.0$\pm$11.0 & 38.5$\pm$16.5 &  39.0$^{+10}_{-1}$ & 32.8$^{+3}_{-1}$ \\           
 & D.M.~[mag] & 7.72$\pm$0.08 & 7.76$\pm$0.19 & 7.77$^{+0.03}_{-0.05}$ & 7.79$^{+0.04}_{-0.07}$ \\
 & (Dist.~[pc]) & (350$\pm$13) & (370$^{+19}_{-43}$) & (358$^{+5}_{-8}$) &  (361$^{+7}_{-11}$) \\
\hline 
BCAH98\tablenotemark{c} & Age~[Myr] & 65.2$\pm$14.1 & 31.5$\pm$10.5  & 68.7$^{+2}_{-12}$ & 39.6$^{+1}_{-5}$ \\
 & D.M.~[mag] & 7.72$\pm$0.05 & 7.94$\pm$ 0.10 & 7.72$^{+0.05}_{-0.01}$ & 7.88$^{+0.11}_{-0.01}$ \\
 & (Dist.~[pc]) &  (350$\pm$8) &  (387$^{+18}_{-17}$) &  (350$^{+8}_{-2}$) &  (377$^{+19}_{-2}$) \\
\hline
SDF00\tablenotemark{c} & Age~[Myr]  & 49.5$\pm$9.5 & 43.0$\pm$16.0 & 46.4$^{+10}_{-3}$ & 27.5$^{+2}_{-1}$ \\
 & D.M.~[mag] & 7.72$\pm$0.10 & 7.86$\pm$0.11 & 7.75$^{+0.04}_{-0.05}$ & 7.96$^{+0.03}_{-0.06}$  \\
 & (Dist.~[pc]) &  (350$\pm$16) &  (373$^{+19}_{-18}$) & (355$^{+6}_{-8}$) &  (391$^{+6}_{-11}$) \\
\enddata
\tablenotetext{a}{Intrinsic distance moduli were measured using an extinction of $A_{V} = 0.54$ mag.}
\tablenotetext{b}{Median of values from all membership catalogs for a given model and color. Uncertainties quoted are the range of possible values given the errors on the individual best-fit isochrone models. We did not include H$\alpha$ isochrone measurement in median and range calculation due to their large uncertainties.}
\tablenotetext{c}{DAM97 -- \citet{DAntona1997} ; BCAH98 -- \citet{Baraffe1998} ; SDF00 -- \citet{Siess2000}}
\end{deluxetable*}

%% file: tab6.tex
\begin{deluxetable*}{l l c c c c c c}
\tablecolumns{8}
\setlength{\tabcolsep}{0.05in}
\tablewidth{0pt}
\tabletypesize{\footnotesize}
\tablecaption{Catalog of IC~4665 Photometric Members\label{tab.isomem}}
\tablehead{
\colhead{Mosaic} & \colhead{Object} & \colhead{V} & \colhead{$B-V$} & \colhead{V-I$_{c}$} & 
\colhead{Phot Mem\tablenotemark{a}} & \colhead{Phot Mem\tablenotemark{a}} & \colhead{Mem Key\tablenotemark{b}}\\
\colhead{Field} & \colhead{Number} & \colhead{} & \colhead{} & \colhead{} &
\colhead{[$B-V$]} & \colhead{[$V-I_{c}$]} & \colhead{}
}
\startdata
F1 & 427 & 13.069$\pm$0.002 &  0.816$\pm$0.005 & 0.997$\pm$0.004 & Y & Y & Li,RV,J, \\
F6 & 347 & 12.729$\pm$0.002 &  0.835$\pm$0.004 & 1.018$\pm$0.003 & N & Y & Li,XR,RP \\
F6 &  59 & 14.475$\pm$0.006 &  1.252$\pm$0.019 & 1.274$\pm$0.009 & Y & Y & RV \\
F6 & 338 & 12.095$\pm$0.001 &  0.628$\pm$0.002 & 0.754$\pm$0.002 & Y & Y & \nodata \\
\enddata
\tablecomments{Only a few rows are displayed here to show the content of the table, the full table is available in electronic form.}
\tablenotetext{a}{Point lies within its photometric uncertainty of the isochrone using the specified color.}
\tablenotetext{b}{Ancillary membership criterion. See Section \ref{sec.lowmass} for source references. Li = Lithium absorption, XR = X-ray emission, H = H$\alpha$ emission, RV = Radial Velocity, RP = detected rapid rotation period, J = listed as member in \citet{Jeffries2009}, and '...' = photometric member only.}
\end{deluxetable*}